%% file: shap1013.tex
\documentclass[preprint]{aastex}
\usepackage{apjfonts}
\usepackage{graphicx}
\usepackage{psfig}
\usepackage{float}

\usepackage{url}

\usepackage{graphics}
\usepackage{longtable}
\usepackage{amssymb,amsmath}
\usepackage{lscape}

\graphicspath{{Figures/}}

\newcommand{\msun}{\ensuremath{\rm{M}_\odot}}

\newcommand{\Lsun}{\ensuremath{\rm{L}_\odot}}

\newcommand{\swift}{{\it Swift}}

\newcommand{\HbFWHM}{\ensuremath{5490 \pm 1140 \rm{ \: km \: s}^{-1}}}
\newcommand{\Hbdisp}{\ensuremath{2200 \pm 230 \rm{ \: km \: s}^{-1}}}
\newcommand{\HeIIFWHM}{\ensuremath{10050 \pm 2590 \rm{ \: km \: s}^{-1}}}
\newcommand{\HeIIdisp}{\ensuremath{4180 \pm 450 \rm{ \: km \: s}^{-1}}}
\newcommand{\OurMbh}{\ensuremath{(4\pm1) \times 10^7} \msun}

\input{xlag_fit.tex}

\begin{document}

\title{The Man Behind the Curtain: 
X-rays Drive the UV through NIR Variability in the 2013 AGN Outburst in NGC~2617}
\shorttitle{NGC~2617}
\shortauthors{Shappee et al.}	

\author{
{B.~J.~Shappee}\altaffilmark{1,2}, 
{J.~L.~Prieto}\altaffilmark{3,4},
{D.~Grupe}\altaffilmark{5},
{C.~S.~Kochanek}\altaffilmark{1,6},
{K.~Z.~Stanek}\altaffilmark{1,6},
{G.~De Rosa}\altaffilmark{1},
{S.~Mathur}\altaffilmark{1,6},
{Y.~Zu}\altaffilmark{1},
{B.~M.~Peterson}\altaffilmark{1,6},
{R.~W.~Pogge}\altaffilmark{1,6},
{S.~Komossa}\altaffilmark{7},
{M.~Im}\altaffilmark{8},
{J.~Jencson}\altaffilmark{1},
{T.W-S.~Holoien}\altaffilmark{1},
{U.~Basu}\altaffilmark{1,9},
{J.~F.~Beacom}\altaffilmark{1,6,10},
{D.~M.~Szczygie{\l}}\altaffilmark{11},
{J.~Brimacombe}\altaffilmark{12},
{S.~Adams}\altaffilmark{1},
{A.~Campillay}\altaffilmark{13},
{C.~Choi}\altaffilmark{8},
{C.~Contreras}\altaffilmark{13},
{M.~Dietrich}\altaffilmark{14},
{M.~Dubberley}\altaffilmark{15},
{M.~Elphick}\altaffilmark{15},
{S.~Foale}\altaffilmark{15},
{M.~Giustini}\altaffilmark{16},
{C.~Gonzalez}\altaffilmark{13},
{E.~Hawkins}\altaffilmark{15},
{D.~A.~Howell}\altaffilmark{15,17},
{E.~Y.~Hsiao}\altaffilmark{13},
{M.~Koss}\altaffilmark{18},
{K.~M.~Leighly}\altaffilmark{19},
{N.~Morrell}\altaffilmark{13},
{D.~Mudd}\altaffilmark{1},
{D.~Mullins}\altaffilmark{15},
{J.~M.~Nugent}\altaffilmark{19},
{J.~Parrent}\altaffilmark{15},
{M.~M.~Phillips}\altaffilmark{13},
{G.~Pojmanski}\altaffilmark{11},
{W.~Rosing}\altaffilmark{15},
{R.~Ross}\altaffilmark{15},
{D.~Sand}\altaffilmark{20},
{D.~M.~Terndrup}\altaffilmark{1},
{S.~Valenti}\altaffilmark{15,17},
{Z.~Walker}\altaffilmark{15},
and 
{Y.~Yoon}\altaffilmark{8}
}

\email{shappee@astronomy.ohio-state.edu}

\altaffiltext{1}{Department of Astronomy, The Ohio State University, Columbus, Ohio 43210, USA}
\altaffiltext{2}{NSF Graduate Fellow}
\altaffiltext{3}{Department of Astrophysical Sciences, Princeton University, Princeton, NJ 08544, USA}
\altaffiltext{4}{Carnegie-Princeton Fellow}
\altaffiltext{5}{Department of Astronomy and Astrophysics, Pennsylvania State University, 525 Davey Lab, University Park,PA 16802, USA}
\altaffiltext{6}{Center for Cosmology and AstroParticle Physics, The Ohio State University, 191 W.\ Woodruff Ave., Columbus, OH 43210, USA}
\altaffiltext{7}{Max-Planck Institut f{\"{u}}r Radioastronomie, Auf dem H\"ugel 69, 53121 Bonn, Germany}
\altaffiltext{8}{CEOU/Department of Physics and Astronomy, Seoul National University, Seoul 151-742, Republic of Korea.}
\altaffiltext{9}{Grove City High School, 4665 Hoover Road, Grove City, OH 43123, USA}
\altaffiltext{10}{Department of Physics, The Ohio State University, 191 W. Woodruff Ave., Columbus, OH 43210, USA}
\altaffiltext{11}{Warsaw University Astronomical Observatory, Al. Ujazdowskie 4, 00-478 Warsaw, Poland}
\altaffiltext{12}{Coral Towers Observatory, Cairns, Queensland 4870, Australia}
\altaffiltext{13}{Carnegie Observatories, Las Campanas Observatory, Colina El Pino, Casilla 601, Chile}
\altaffiltext{14}{Department of Physics and Astronomy, Ohio University, 251B Clippinger Labs, Athens, OH 45701, USA}
\altaffiltext{15}{Las Cumbres Observatory Global Telescope Network, 6740 Cortona Drive, Suite 102, Santa Barbara, CA 93117, USA}
\altaffiltext{16}{XMM-Newton Science Operation Centre, European Space Astronomy Centre (ESAC)/ESA, Villanueva de la Ca\~nada, E-28691 Madrid, Spain}
\altaffiltext{17}{Department of Physics, University of California, Santa Barbara, CA 93106, USA}
\altaffiltext{18}{Institute for Astronomy, University of Hawaii, 2680 Woodlawn Drive, Honolulu, HI 96822, USA}
\altaffiltext{19}{Homer L. Dodge Department of Physics and Astronomy, The University of Oklahoma, 440 W. Brooks St., Norman, OK 73019, USA}
\altaffiltext{20}{Texas Tech University, Physics Department, Box 41051, Lubbock, TX 79409-1051, USA}

\date{\today}

\def\Civ{C\,{\sc iv}}
\def\OIV{O\,{\sc iv}}
\def\Mgii{Mg\,{\sc ii}}
\def\Feii{Fe\,{\sc ii}}
\def\Feiii{Fe\,{\sc iii}}
\def\NV{N\,{\sc v}}
\def\Lya{Ly$\alpha$}
\def\Lyb{Ly$\beta$}
\def\SiII{Si\,{\rm {\sc ii}}}
\def\OI{O\,{\sc i}+Si\,{\sc ii}}
\def\CII{C\,{\sc ii}}
\def\CIIforb{[C\,{\sc ii}]}
\def\CIII{C\,{\sc iii}]}
\def\SiIV{Si\,{\sc iv}}
\def\HII{H\,{\sc ii}}
\def\HI{H\,{\sc i}}
\def\Hb{H$\beta$}
\def\Ha{H$\alpha$}
\def\OIII{[O\,{\sc iii}]}
\def\HeII{He\,{\sc ii}}
\def\lsim{\mathrel{\rlap{\lower 3pt \hbox{$\sim$}} \raise 2.0pt \hbox{$<$}}}
\def\gsim{\mathrel{\rlap{\lower 3pt \hbox{$\sim$}} \raise 2.0pt \hbox{$>$}}}
\def\sline{$\sigma_{\rm line}$}
\def\Msun{M$_\odot$}
\def\Lsun{L$_\odot$}
\def\Mbh{$M_{\rm BH}$}
\def\lLl{$\lambda L_{\lambda}$}

\begin{abstract}

After the All-Sky Automated Survey for SuperNovae (ASAS-SN) discovered a significant brightening of the inner region of NGC~2617, we began a $\sim 70$ day photometric and spectroscopic monitoring campaign from the X-ray through near-infrared (NIR) wavelengths.  We report that NGC~2617 went through a dramatic outburst, during which its X-ray flux increased by over an order of magnitude followed by an increase of its optical/ultraviolet (UV) continuum flux by almost an order of magnitude. NGC~2617, classified as a Seyfert 1.8 galaxy in 2003, is now a Seyfert 1 due to the appearance of broad optical emission lines and a continuum blue bump. Such ``changing look Active Galactic Nuclei (AGN)'' are rare and provide us with important insights about AGN physics.  Based on the H$\beta$ line width and the radius-luminosity relation, we estimate the mass of central black hole to be \OurMbh.  When we cross-correlate the light curves, we find that the disk emission lags the X-rays, with the lag becoming longer as we move from the UV ($2\--3$ days) to the NIR ($6\--9$ days).  Also, the NIR is more heavily temporally smoothed than the UV.  This can largely be explained by a simple model of a thermally emitting thin disk around a black hole of the estimated mass that is illuminated by the observed, variable X-ray fluxes.

\end{abstract}

\section{Introduction}
\label{sec:introduction}

NGC~2617 is a face-on Sc galaxy at a redshift of $z=0.0142$ \citep{paturel03}, corresponding to a Hubble flow corrected distance of $64$ Mpc. \citet{moran96} classified it as a Seyfert 1.8 galaxy. The galaxy displayed only narrow emission lines except for a weak, broad base to the H$\alpha$ line.  It lacked both the broad emission lines that are characteristic of Seyfert 1 galaxies, and a bright power-law continuum.  Here we report on the dramatic transformation of NGC~2617 into a Seyfert 1 and the monitoring of the (presumably) associated X-ray--NIR outburst.

It is well known that both the broad line and continuum components of active galactic nuclei (AGN) are time variable at all wavelengths.  This enables, for example, the reverberation mapping (RM) technique for studying black hole (BH) masses (e.g., \citealp{peterson04, grier12a, grier12b, grier13a}) or measuring time delays in gravitational lenses (e.g. \citealp{tewes13}).  However, there are only a few cases where the changes are so large that the broad-line components of H$\alpha$ and H$\beta$ appear in a Seyfert 2 or vanish from a Seyfert 1.  Vanishing broad-line components have been reported for NGC~7603 \citep{tohline76}, NGC~4151 \citep{penston84}, Mrk~372 \citep{gregory91}, and 3C390.3 \citep{penston84, veilleux91}, and broad-line components have appeared in Mrk~6 \citep{khachikian71}, Mrk~1018 \citep{cohen86}, Mrk~993 \citep{tran92}, NGC~1097 \citep{storchibergmann93}, and NGC~7582 \citep{aretxaga99}. ``Changing-look'' AGN have been observed in X-rays as well (e.g., \citealp{matt03}). The X-ray absorption in IRAS $09104+4109$, known as a Seyfert 2 AGN, changed from being Compton-thick to Compton-thin, making it essentially a Seyfert 1 AGN (e.g., \citealp{lusso13}). Similar variability was subsequently observed in a several other AGN \citep{marchese12}.  These strong variations must be telling us something fundamental about the structure of the broad-line region or the obscuring material.

Additionally, some AGN show large changes in their (X-ray) continuum emission. An example is IC 3599, which flared by approximately a factor of a hundred in soft X-rays, accompanied by the appearance and then disappearance of broad and narrow optical emission lines \citep{brandt95, grupe95a, komossa99}.  Other AGN show high-amplitude X-ray variability without significant emission line variability. Among the most extreme cases are WPVS007 (e.g., \citealp{grupe95b, grupe13}), E 1615+061 \citep{piro97}, Mkn 335 \citep{grupe07, grupe08, grupe12}, and  PHL 1092 \citep{miniutti09}.  There are examples of still more extreme changes in X-ray emission, up to factors of several thousand, which have been detected in optically quiescent galaxies.  These events have been interpreted as stellar tidal disruption flares, which are thought to occur when stars are disrupted and subsequently accreted by SMBHs in galaxy cores (see \citealp{komossa12} for a recent review).

Broad emission lines are one of the defining characteristics of Seyfert 1 AGN. In the unification model of AGN, it is believed that Seyfert 2 AGN are Seyfert 1 AGN where the broad emission line region (BLR) is being blocked from view by an obscuring torus of gas and dust.  However, the intermediate Seyfert types (1.8 and 1.9) do not easily fit into this picture.  \citet{goodrich89} suggested that the difference between Seyfert 1 AGN and these intermediate Seyfert types is simply that the intermediate Seyferts are viewed through a screen of dust and that the variability of these lines is due to the changing optical depth of this material.  Beyond viewing angles and obscuration, the fundamental parameters determining the presence of a BLR are still uncertain, but are  likely related to the BH mass and accretion rate. For example, \citet{chakravorty14} have proposed that the presence of a BLR depends on the BH mass, with objects below a critical mass showing no BLR. \citet{nicastro00} offers a different model, arguing that the accretion rate relative to Eddington is the fundamental parameter. In this model, the inner radius of the BLR is the transition radius at which the accretion disk goes from being radiation-to-gas pressure dominated and below a critical mass accretion rate no BLR should form.  \citet{czerny11} offer a third model. They note that the accretion disk temperature at the radius of the BLR is about 1000 K in all AGN. This is close to the dust sublimation temperature, which may allow the formation of a radiation-driven wind which then forms the BLR.  

On the other hand, the origins of the optical/UV AGN continuum are known to be thermal emission from an accretion disk around a supermassive BH \citep{shields78}. The exact structure of the disk is uncertain, and disk size measurements using gravitational microlensing generally find sizes larger than expected from simple thin disk theory (e.g., \citealp{morgan10}).  

The nature of the X-ray continuum and its relation to the UV continuum, however, is still a matter of debate.  The X-rays are generally believed to arise from Compton up-scattering of disk photons in a hot corona (e.g., \citealp{reynolds03}).  Alternatively, optical/UV continuum may be produced from down-scattering of X-ray photons (e.g., \citealp{cameron12} and references therein).    Theoretical models have postulated a broad range of physical sizes for this hot corona. However, recent X-ray gravitational microlensing results consistently find sizes comparable to the inner edge of the accretion disk (e.g., \citealp{morgan08, mosquera13}), thus ruling out models with very extended coronae.  

The causal ordering of X-ray and disk variability is, however, unclear, since attempts to measure time lags between these emission components have generally yielded only null results or tentative detections (e.g. \citealp{breedt09, cameron12}).  These lags, however, could be a powerful tool to distinguish between the two physical processes most likely responsible for strong X-ray--NIR variability observed in NGC 2617 which are 1) locally-generated (e.g. viscous) perturbations propagating inward through the accretion disk or 2) an increase in the X-ray flux coming from the smaller, central hot corona which heats the inside of the disk first and moves outward.  Thus, changes in accretion rate might be expected to produce ``outside-in'' variations, moving from red to blue to X-rays, while changes in coronal emission might be expected to produce ``inside-out'' variations with the increased X-ray flux irradiating the disk and driving increased blue and then red emission. 

Delays between wavelengths might also be measurable for the UV/optical emission.  The best case to date is that of NGC 7469 \citep{wanders97, collier98, peterson98, kriss00}, where longer wavelengths increasingly lag behind those in the 1315 \AA{} continuum. The lag detections are significant only at about the 2-sigma level, however. Surprisingly, the X-ray variations \citep{nandra98} did not correlate particularly well with those in the UV/optical, and in retrospect this is probably attributable to variable absorption of the X-ray spectrum (e.g., \citealp{risaliti11}). Lags between shorter- and longer-wavelength continuum variations have been detected in several sources (e.g., \citealp{sergeev05, cackett07}) at marginal significance, and always in the sense that the short-wavelength variations lead those at longer wavelengths.

The dramatic state change and outburst we observe in NGC~2617 provides a laboratory for exploring the relations between these AGN components.  In \S\ref{sec:Obs} we describe our NIR $\--$ X-ray observations during the outburst of NGC~2617.  In \S\ref{sec:Anal} we search for variations in the broad H$\beta$ and \HeII$\lambda 4686$ lines, estimate the central BH mass, and cross-correlate the X-ray to UV--NIR light curves to measure the time delays between the variability in different bands.  We present a simple physical model that reproduces the observed UV--IR variability in \S\ref{sec:Model} and summarize our results in \S\ref{sec:SumTotal}. Throughout the paper all dates are given in UT.

\section{Observations}
\label{sec:Obs}

We have been conducting the All-Sky Automated Survey for SuperNovae (ASAS-SN\footnote{\url{http://www.astronomy.ohio-state.edu/~assassin}}  or ``Assassin''), which scans the extragalactic sky visible from Hawaii roughly once every five nights in the $V$-band.  On 2013  Apr. 10.27, a transient source alert was triggered by the brightening of the central region of NGC~2617. Follow-up observations confirmed that the nucleus was the source of the outburst and a follow-up spectrum showed that the source now has the strong broad emission lines and continuum shape characteristic of a Seyfert 1. Given the strong outburst and dramatic change in the spectral properties, we requested and obtained \swift\ \citep{gehrels04} target of opportunity (ToO) observations to monitor NGC~2617 in the optical, UV, and X-rays, which found that NGC~2617 continued to brighten. During this outburst we were able to obtain an extensive time series of optical spectroscopy and NIR/optical/UV/X-ray photometry that are presented in this paper. The INTErnational Gamma-Ray Astrophysics Laboratory (INTEGRAL; \citealp{winkler03}) also detected the outburst in hard X-rays ($17\--60$~keV), as has the European VLBI Network real-time Interferometer (e-VLBI) and the Very Long Baseline Array (VLBA) in the radio at 1.6 GHz and 5 and 1.7 GHz, respectively.  Finally, XMM-Newton observations were obtained for NGC~2617 during the outburst, and these observations will be discussed in a companion paper (Giustini et al. in prep.).  In the following subsections we describe the observations of NGC~2617 in detail.

\subsection{ASAS-SN Discovery of NGC~2617 Outburst}
\label{sec:ASAS}

ASAS-SN is a long-term project to monitor the whole extragalactic sky to find nearby supernovae (SNe) and other transient sources.  We began running our real-time search for variable sources in April 2013 with our first unit, Brutus. Brutus presently consists of two telescopes on a common mount which is hosted by Las Cumbres Observatory Global Telescope Network (LCOGT; \citealp{brown13}) in the Faulkes Telescope North enclosure on Mount Haleakala, Hawaii.  Each telescope consists of a 14-cm aperture Nikon telephoto lens and an FLI ProLine CCD camera with a Fairchild Imaging 2k $\times$ 2k thinned CCD, giving a $4.47\times4.47$ degree field-of-view and a 7$^{''}\!\!\!\!.8$ pixel scale.  The cameras are mounted to have moderately overlapping fields-of-view.  On a typical clear night, Brutus can survey more than 5000~deg$^2$, and with the effects of weather we can survey the extragalactic sky visible from Hawaii roughly once every 5 days.  The data are reduced in real-time, and we can search for transient candidates about an hour after the data are taken using an automated difference imaging pipeline.  We are now meeting, and frequently exceeding, our current depth goal of $V \approx 16$~mag, corresponding to the apparent brightness at maximum light of core-collapse SNe within 30~Mpc and SNe Ia out to 100~Mpc.

\begin{figure}
	\centerline{
		\includegraphics[width=17cm]{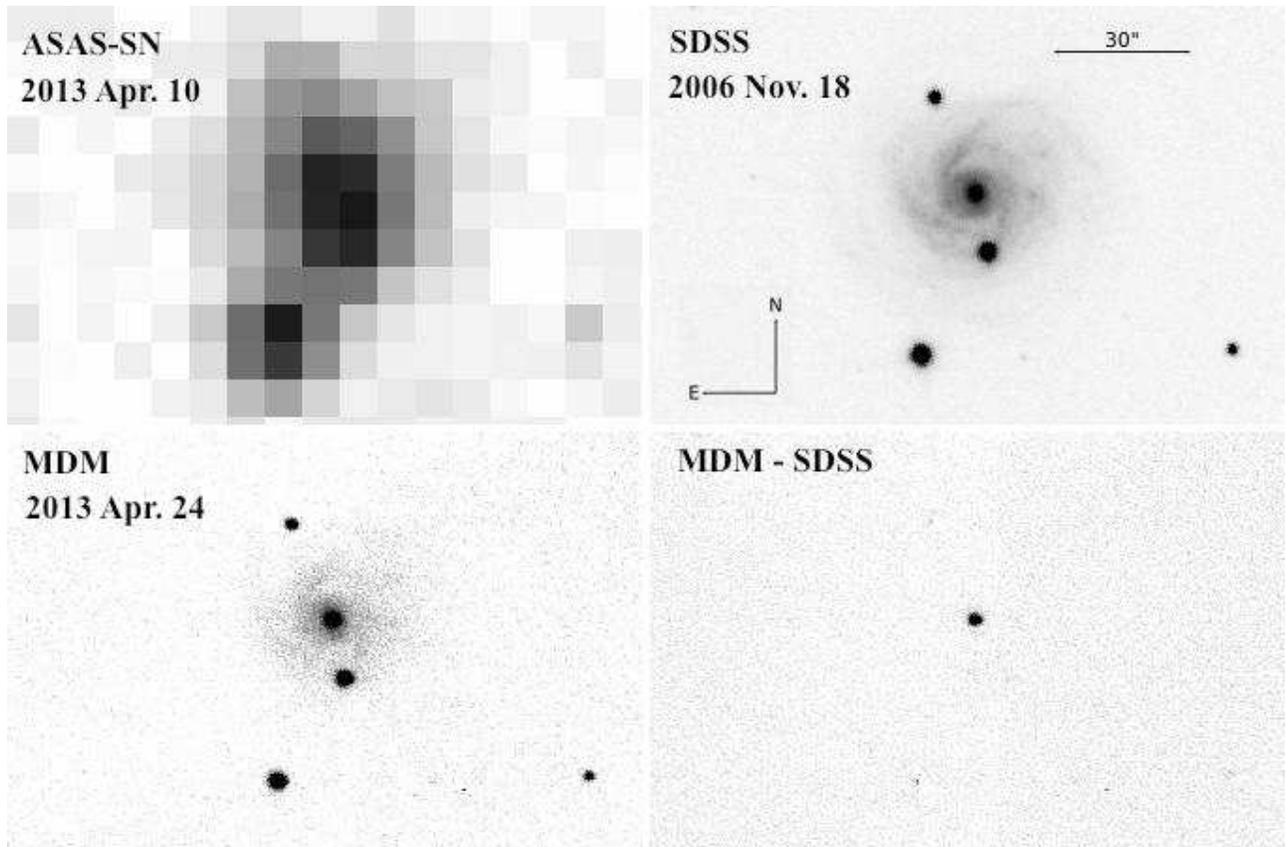}
	}
	\caption{Comparison between our ASAS-SN discovery image (top-left), the archival SDSS image (top-right), our follow-up MDM image (bottom-left), and a difference image showing the brightening of the central region between the SDSS and MDM images (bottom-right). All images are on the same angular scale. This increase in flux is associated with the center of the galaxy to within 0$^{''}\!\!\!\!.03$ or $\sim9$ pc at NGC~2617.}
	\label{fig:ASASdisc}
\end{figure}

Our transient source detection pipeline was triggered by an image taken at 2013  Apr. 10.27, detecting a $\sim 10\%$ flux increase from the inner region of NGC~2617 (Figure~\ref{fig:ASASdisc}), equivalent to detecting a new point source source of $V \sim16.7$ mag superimposed on the image of the galaxy \citep{shappee13ATELa}. It seemed likely to be AGN variability, but the AGN and the host are not separately resolved given ASAS-SN's 16$^{''}$ resolution.  We obtained a follow-up image on Apr. 24.14 with the Ohio State Multi-Object Spectrograph (OSMOS; \citealp{martini11}) at the 2.4-m telescope at MDM Observatory, which showed that the central region of NGC~2617 continued to brighten and was a factor of $\sim3.3$ brighter than the archival Sloan Digital Sky Survey (SDSS; \citealp{york00}) Data Release 9 \citep{ahn12} $g$-band image, as shown in Figure~\ref{fig:ASASdisc}. This increase in flux was within 0$^{''}\!\!\!\!.03$ of the center of the galaxy, corresponding to a projected physical distance of less 
than $\sim9$ pc at NGC~2617.

\begin{figure}
	\centerline{
		\includegraphics[width=17cm]{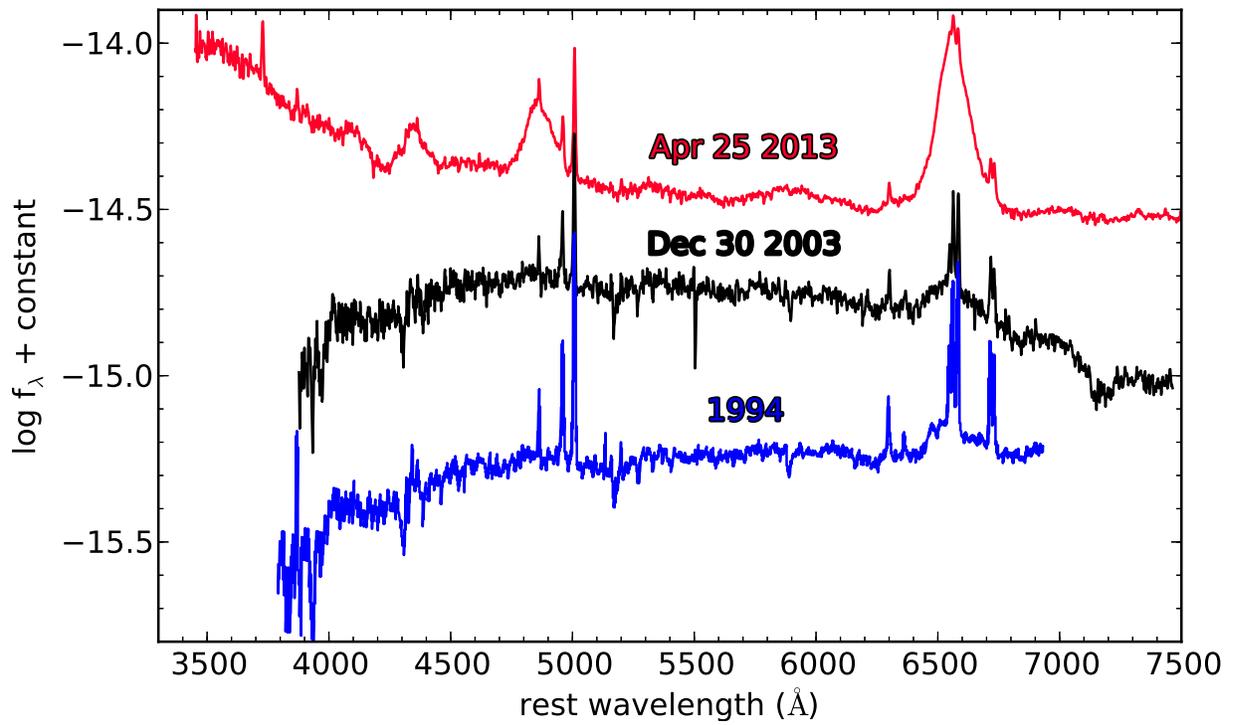}
	}
	\caption{Comparison between the APO spectrum from 2013 April 25 (top, red), the 6dFGS spectrum from 2003 Dec. 30 (middle, black), and the \citet{moran96} spectrum from 1994 (bottom, blue).  Note that the absolute flux calibration is arbitrary.  NGC~2617 underwent a dramatic spectral evolution changing from a Seyfert 1.8 to a Seyfert 1 sometime between 2003 and 2013. }
	\label{fig:spec_chg}
\end{figure}

We then requested director's discretionary (DD) time on the Apache Point Observatory (APO) 3.5-m telescope and obtained an optical spectrum (range $3500\--9600$ \AA) using the Dual Imaging Spectrograph (DIS) on Apr. 25.1.  Figure~\ref{fig:spec_chg} shows a comparison between this spectrum, a spectrum from the 6dF Galaxy Survey (6dFGS) \citep{jones04, jones09} taken on 2003 Apr. 25, and the spectrum from \citet{moran96} taken in 1994. This figure illustrates the dramatic spectral changes of NGC~2617. Along with the increase in the continuum flux, NGC~2617 changed spectral type from a Seyfert 1.8 to a Seyfert 1, with the appearance of strong broad-line emission in H$\alpha$ and H$\beta$.  The spectrum also shows the appearance of a strong ``blue bump'' shortward of 4000 \AA. It is possible that the spectral changes observed in NGC~2617 are not correlated with the large AGN outburst discovered by ASAS-SN. However, given the rarity of both ``changing-look'' AGN and powerful X-ray flares, this seems a remote possibility even if the available data cannot prove an association.

\subsection{Swift ToO Optical, Ultraviolet, and X-ray Observations}
\label{sec:SWIFT}

The first 2 ks Swift X-ray Telescope (XRT; \citealp{burrows05}) observation obtained on 2013  Apr. 30.93 detected NGC~2617 with a flux of $3.1 \times 10^{-11}$ ergs cm$^{-2}$ s$^{-1}$ (0.3--10~keV; \citealp{mathur13ATELa}).  This X-ray flux is significantly higher than the fluxes measured for NGC~2617 by both the \textit{XMM-Newton} slew survey in 2007 \citep{saxton08} of $9.0 \times 10^{-12}$ ergs cm$^{-2}$ s$^{-1}$  (0.212 keV) and the  R{\"o}ntgen Satellite (ROSAT; \citealp{truemper82}) of $(6.9 \pm 1.0) \times 10^{-12}$ ergs cm$^{-2}$ s$^{-1}$ (0.1--2.4~keV; \citealp{boller92}). Based on this initial result we requested and were granted further \swift\ ToO observations. By 2013  May 8.04, continued monitoring with \swift\ XRT showed an increase of another factor of four in the X-ray flux, so the AGN was now an order of magnitude brighter than during the \textit{XMM-Newton} Slew Survey observation in 2007.  Parallel optical and UV observations obtained with \swift's UltraViolet/Optical Telescope (UVOT; \citealp{roming05}) also showed an increase in flux, but on a longer timescale \citep{shappee13ATELb}. In total, we obtained \swift\ XRT and UVOT observations spanning $\sim 50$ days with an almost daily cadence from 2013 May 01 through June 20 when it became unobservable due to \swift's Sun-constraint.

\subsubsection{\swift\ UVOT Observations}
\label{sec:UVOT}

\begin{figure*}
	\centerline{
		\includegraphics[height=17cm]{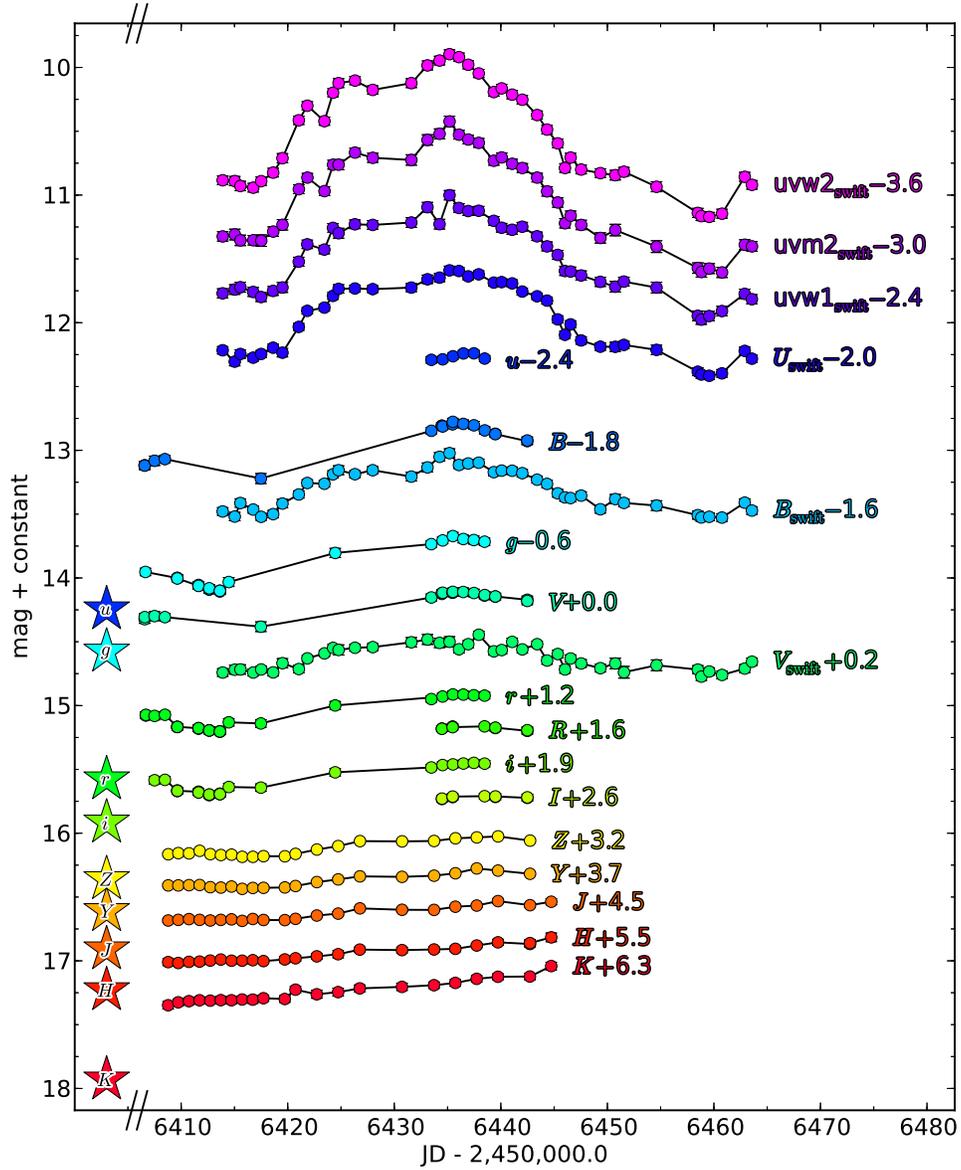}
	}
	\caption{NIR--UV photometric observations of NGC~2617 spanning $\sim60$ days from 2013 Apr. 24 through June 20. The filled circles are our new observations while the large stars to the left are archival magnitudes from SDSS and 2MASS.  Error bars are shown, but they are usually smaller than the points.}
	\label{fig:phot}
\end{figure*}

The \swift\ UVOT observations of NGC~2617 were obtained in 6 filters \citep{poole08}: $V$ (5469~\AA), $B$ (4392~\AA), $U$ (3465~\AA), $UVW1$ (2600~\AA), $UVM2$ (2246~\AA), and $UVW2$ (1928~\AA). We used the UVOT software task  {\it uvotsource} to extract the source and background counts from a 5$^{''}\!\!\!\!.0$ radius region and a 5$^{''}\!\!\!\!.0$ to 20$^{''}\!\!\!\!.0$ annulus, respectively. The UVOT count rates were converted into magnitudes and fluxes  based on the most recent UVOT calibration \citep{poole08, breeveld10}. The UVOT UV/optical photometry is reported in Table~\ref{tab:phot} and shown in Figure~\ref{fig:phot}. 

\input{tabphot.tex}

\subsubsection{\swift\ XRT Observations}
\label{sec:XRT}

\begin{figure}
	\centerline{
		\includegraphics[height=13cm]{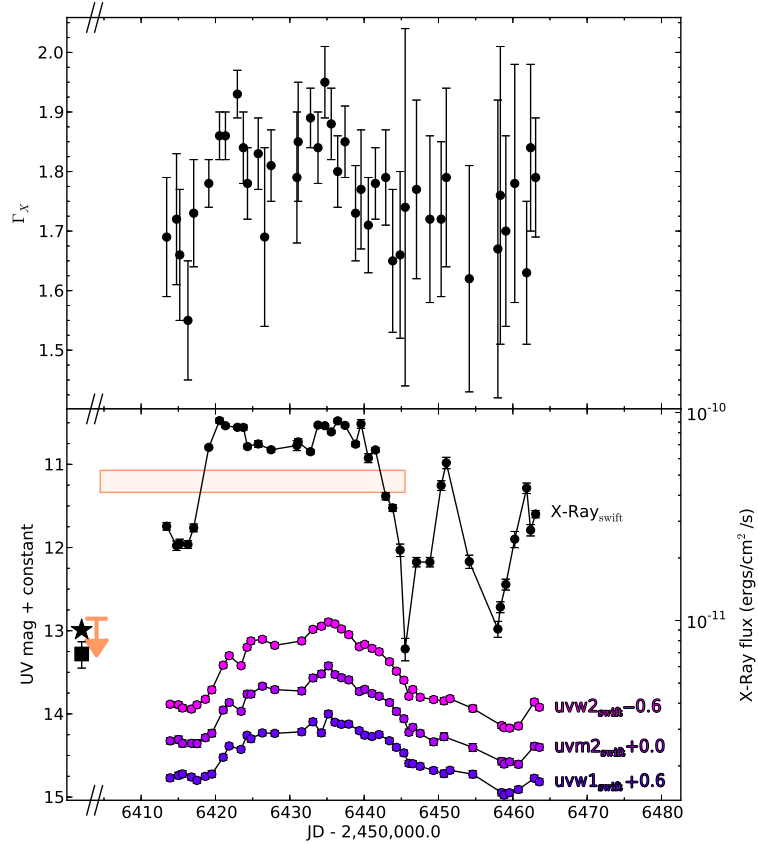}
	}
	\caption{{\it Lower Panel}: The \swift\ XRT $0.3\--10$ keV X-ray (black circles), \swift\ UVOT UV (magenta, violet, and purple circles), and INTEGRAL IBIS $17\--60$~keV X-ray (apricot rectangle) observation(s) of NGC~2617.  {\it Upper Panel}: The \swift\ XRT X-ray photon spectral index ($\Gamma$). The apricot rectangle represents the INTEGRAL IBIS hard X-ray flux measurement, where the width and height represent the span of the observation and the measurement uncertainty, respectively. The large star and square to the left are archival \textit{XMM-Newton} measurement (0.2--12 keV) from 2007 and archival ROSAT measurement (0.1--2.4 keV) from 1990-1991, respectively. The apricot upper limit shown to the left is an archival flux limit from the 70-month \swift/BAT survey (14--195 keV; see \S\ref{sec:HardX}). As NGC~2617 brightened the X-ray spectrum softened. Both the UV and X-ray light curves show a similar double humped feature during the outburst, but with a 2-3 day delay (see \S\ref{sec:timed}). }
	\label{fig:xray}
\end{figure}

The XRT was usually operating in Windowed Timing (WT) mode \citep{hill04} but it was in photon counting (PC) mode for some observations at the beginning and middle of the monitoring campaign.  The data were reduced by the task {\it xrtpipeline} version 0.12.6, which is included in the HEASOFT package 6.12. For the PC mode data the source counts were collected in a circle with a radius of 70$^{''}\!\!\!\!.7$. Background counts were estimated from a source-free region 212$^{''}$ in radius using the task {\it xselect} (version 2.4b). The WT source and background data were selected in a box with a length of 40 pixels.  Auxiliary response files were created using the XRT task {\it xrtmkarf}. The spectra were rebinned with 20 counts per bin using the task {\it grppha}. We applied the response files {\it swxpc0to12s6\_20010101v013.rmf} and {\it swxwt0to2s6\_20070901v012.rmf} for the PC and WT data, respectively. The re-binned $0.3\--10$~keV  spectra were modeled with XSPEC v.12.7 using a single power law and Galactic absorption corresponding to a hydrogen column density of $N_{\rm{H}}$ = 3.64$\times$10$^{20}$~cm$^{-2}$ \citep{kalberla05}. The \swift\ XRT X-ray fluxes and XRT X-ray photon spectral indices ($\Gamma$) are also reported in Table~\ref{tab:phot}. Figure~\ref{fig:xray} shows both the X-ray flux and UV magnitudes on the same relative scale as well as the X-ray spectral index.  The X-ray spectral index becomes softer as the X-ray flux increases, as has been seen in other AGN (e.g., \citealp{romano02, grupe12}).

\subsection{Additional Photometric Follow-up}
\label{sec:phot}
We obtained optical and NIR photometry from a number of ground-based observatories.  In an attempt to homogenize the photometric data, we performed photometry using a 5$^{''}$ radius aperture on all our data to minimize the differences between the \swift\ UVOT observations, which have a large pixel scale, and our ground-based observations, which have a much smaller pixel scale.  We measured the background within a annulus of 30$^{''}\!\!\!\!.0$ to 60$^{''}\!\!\!\!.0$.  

The $Z$, $Y$, $J$, $H$, and $K$-$\textrm{band}$ observations were obtained with the 3.8-m United Kingdom Infrared Telescope (UKIRT).  The data are calibrated using Two Micron All-Sky Survey (2MASS; \citealp{skrutskie06}) stars within 7$^{'}$ from NGC~2617 and then converted to the WFCAM system following \citet{hodgkin09}.

$B$, $V$, $R$, and $I$-$\textrm{band}$ observations were taken at the LCOGT 1-m facilities in Cerro Tololo, Chile and Sutherland, South Africa \citep{brown13}. All images were reduced following standard procedures and cosmic rays were removed using LACOSMIC \citep{vandokkum01}.  The photometric measurements include a small aperture-corrections ($\lesssim0.02$ mag). The data were calibrated using SDSS stars within 1$^{'}\!\!\!.$5 of NGC~2617 and transformed onto the Johnson-Cousins magnitude system using procedures described by Lupton (2005).\footnote{Lupton (2005); \\ \url{http://www.sdss.org/dr5/algorithms/sdssUBVRITransform.html}}

SDSS $g$, $r$, and $i$-$\textrm{band}$ observations were obtained with the Retractable Optical Camera (RETROCAM; \citealp{morgan05}) and SDSS $g$ and $r$-$\textrm{band}$ observations were obtained with OSMOS both on the MDM 2.4-m telescope.  Small aperture-corrections ($\lesssim0.02$ mag) were applied to the aperture magnitudes, which were then calibrated using SDSS stars within 1$^{'}\!\!\!.$5 of NGC~2617.

$B$, $V$, $u$, $g$, $r$, and $i$-$\textrm{band}$ observations were obtained with the Direct CCD Camera (CCD) at the Swope 1m telescope at Las Campanas Observatory.  Small aperture-corrections ($\lesssim0.02$ mag) were applied to the aperture magnitudes, which were then calibrated using SDSS stars within 1$^{'}\!\!\!.$5 of NGC~2617 and transformed onto the Johnson magnitude system as described by Lupton (2005).

Figures \ref{fig:phot} and \ref{fig:xray} show that the amplitude of the flare was largest in the X-rays, followed by the UV, and then the optical/NIR.  We note, however, that there is significant host galaxy contamination at longer wavelengths which is discussed in \S\ref{sec:sed}. For comparison, we also show the archival SDSS and 2MASS magnitudes of NGC~2617 from 2006  Nov 18 and 1998  Nov 27, respectively.  The 2MASS magnitudes were translated onto the WFCAM system following \citet{hodgkin09}.

\subsection{Spectroscopic Follow-up}
\label{sec:spectra}

We obtained several low-resolution optical spectra of NGC~2617 between 2013 April and June using different telescopes, instruments, and spectroscopic setups. Table~\ref{tab:spec} shows a summary of all the spectra, including date (UT+JD), telescope/instrument, wavelength range, spectral resolution, slit width, seeing estimate, slit position angle, airmass at the beginning of the observation, and exposure time. 

The single slit spectra from DIS at the APO 3.5-m, Boller \& Chivens CCD Spectrograph (CCDS) at the MDM 2.4-m and 1.3-m, OSMOS at the MDM 2.4-m, and the Inamori-Magellan Areal Camera \& Spectrograph (IMACS; \citealp{dressler11}) at the Magellan~6.5-m were all reduced with standard routines in the IRAF {\tt twodspec} and {\tt onedspec} packages. The reductions included bias subtraction, flat-fielding, 1D spectral extraction, wavelength calibration using an arc-lamp, and flux calibration using a spectroscopic standard usually taken the same night. 

The integral field unit spectrum obtained with the Supernova Integral Field Spectrograph (SNIFS; \citealp{lantz04}) at the University of Hawaii (UH) 2.2-m was reduced and extracted using the SNIFS Data Reduction Package \citep{aldering06}. A 3$^{''}\!\!\!\!.$2 aperture was used to extract the final 1D spectrum.

The spectrum obtained with the FLOYDS Spectrograph (Sand et al., in preparation) at the robotic 2-m Faulkes Telescope South (FTS; \citealp{brown13}) was  reduced by an automated reduction pipeline that performs order rectification, flat-fielding, wavelength and flux calibration, and object extraction.  Direct checks of the wavelength solution are also performed using telluric absorption features and night sky lines.

\begin{figure*}
	\centerline{
		\includegraphics[width=13cm]{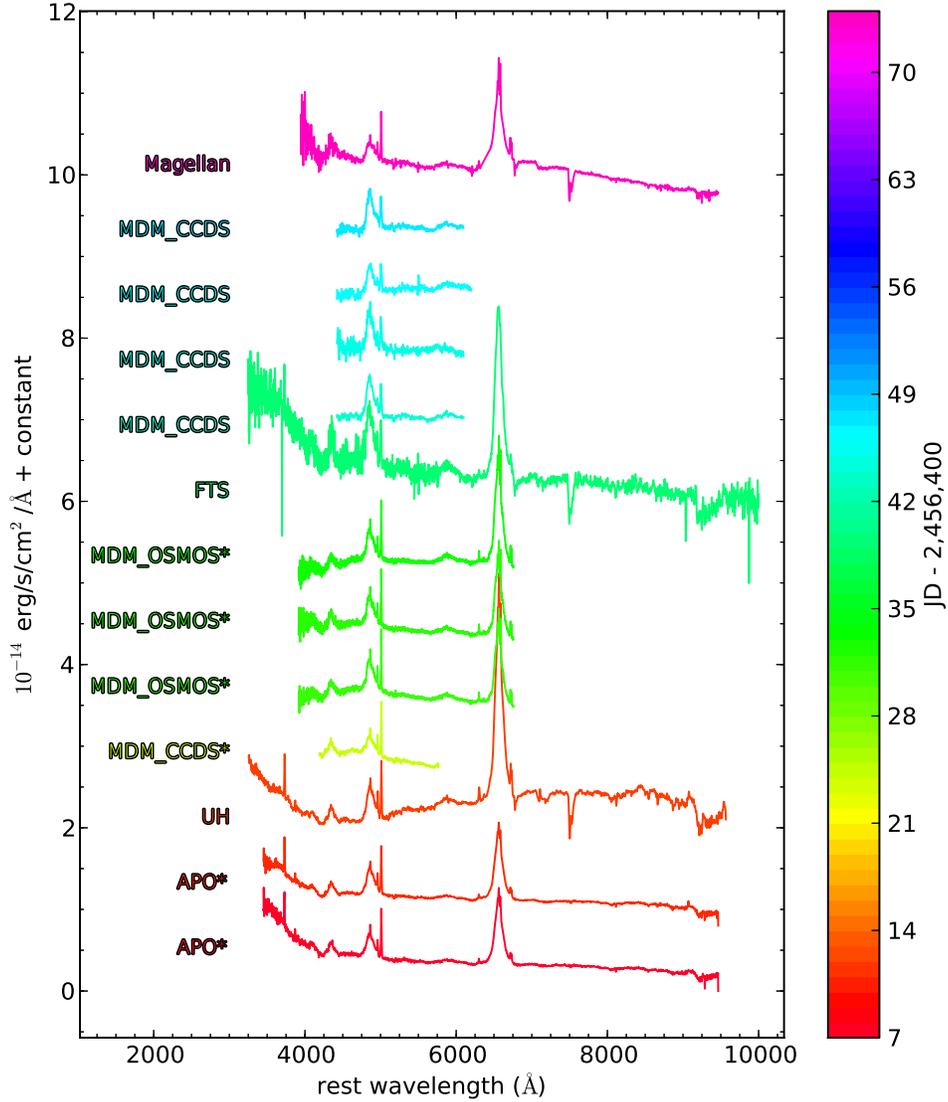}
	}
	\caption{Spectroscopic observations of NGC~2617 spanning $\sim 70$ days from 2013 Apr. 25 -- Jun. 30, 2013. Details of the observations are giving in Table~\ref{tab:spec}. Individual spectra are calibrated by the flux in their narrow \OIII$\lambda$5007 lines. Spectra marked with an asterisk are those employed in the spectral analysis presented in \S\ref{sec:SpecAnal}. The UH spectrum, excluded from our analysis, has some calibration problems leading to distorted line ratios.}
	\label{fig:optical_spec}
\end{figure*}

The spectra were taken under various conditions, many of which were non-photometric.  Additionally, many of the spectra were taken at high airmass and/or during astronomical twilight because the object was close to setting for the season.  These conditions caused the normal flux calibration of the spectra to be poor. We perform an additional step and calibrated each spectrum onto an absolute flux scale under the assumption that the flux of the \OIII$\lambda$5007 narrow emission line is constant as discussed in \S\ref{sec:SpecAnal}. We chose the MDM spectrum taken on 2013  May 12.13 as a reference spectrum for the absolute flux calibration, since it was obtained under clear conditions.  In this spectrum the line flux is $F($\OIII$\lambda 5007)= 5.5 \times 10^{-14}$ erg cm$^{-2}$ s$^{-1}$. These calibrated spectra are shown in Figure~\ref{fig:optical_spec} and our search for temporal variations in the H$\beta$ and \HeII$\lambda 4686$ lines is described in \S\ref{sec:SpecAnal}.  Note, the spectra have not been 
corrected for telluric absorption, but the spectral region near H$\beta$ is not significantly affected.

\input{tabspec.tex}

\subsection{Hard X-Rays}
\label{sec:HardX}

Between 2013  Apr. 22 and 2013  Jun. 2, INTEGRAL discovered a new source, IGR J08357$-$0404, coincident with NGC~2617 during their Galactic Latitude Scan observations of the Puppis region \citep{tsygankov13ATELa}. They detected this source at a 9-sigma significance with a flux of $3.3\pm0.4$ mCrab ($4.7 \times 10^{-11}$ erg cm$^{-2}$ s$^{-1}$) in the $17\--60$~keV energy range. Correcting for the distance to NGC~2617, this flux corresponds to a luminosity of $L=1.86 \times 10^{43}$~erg s$^{-1}$. This observation is shown in Figure~\ref{fig:xray} by an apricot rectangle where the width represents the span of the observation and the height shows its uncertainty. 

\citet{tsygankov13ATELa} also report that this source was undetected, with a 1-sigma upper limit of 1.6~mCrab ($2.3 \times 10^{-11}$ erg cm$^{-2}$ s$^{-1}$), in the INTEGRAL/Imager on-Board the INTEGRAL Satellite (IBIS; \citealp{ubertini03}) nine-year Galactic plane survey (\citealp{krivonos12}).  Correcting for the distance to NGC~2617, this flux upper limit corresponds to a upper limit on the luminosity of $L<9.0 \times 10^{42}$~erg s$^{-1}$. 

Finally, this source is also not present in the 70-month \swift/Burst Alert Telescope (BAT; \citealp{barthelmy05}) survey \citep{baumgartner13}. Conservatively scaling from Figure~1 in \citet{baumgartner13} for the position of NGC 2617, we see that from Dec. 2004 -- Sept. 2010 the exposure time by the 70-month \swift/BAT survey was $> 10$ megaseconds.  Then from Figure~11 in \citet{baumgartner13} we see that this exposure time leads to a 5-sigma upper limit on the hard X-ray flux (14--195 keV) of $< 0.6$~mCrab ($8.5 \times 10^{-12}$ erg cm$^{-2}$ s$^{-1}$).  This upper limit is shown on Figure~\ref{fig:xray}.  Correcting for the distance to NGC~2617, this flux upper limit corresponds to a upper limit on the luminosity of $L < 3.4 \times 10^{42}$~erg s$^{-1}$.

\subsection{Radio}
\label{sec:Radio}

On 2013  Jun. 7, \citet{yang13ATELa} obtained a 7-hour European VLBI Network ToO  observation of NGC 2617 at 1.6~GHz and detected a coincident compact source with a flux density of $1.5\pm0.3$~mJy, corresponding to a radio luminosity of $L_{1.6}=4.8 \times 10^{37}$~erg s$^{-1}$.  From these observations, \citet{yang13ATELa} constrain the source size to be less than 4 milliarcseconds, corresponding to a physical size of $<1.2$~pc at NGC~2617.

On 2013  Jun. 29, \citet{jencson13ATEL} conducted VLBA observations of the nucleus of NGC~2617 at 5 and 1.7 GHz. We detected radio emission with flux densities of $1.5 \pm 0.1$ and $1.6 \pm 0.3$ mJy at 5 and 1.7 GHz, respectively. These measurements indicate little to no variability since the earlier e-EVN observations \citep{yang13ATELa}, and are consistent with a flat spectrum for the nucleus of NGC 2617. \citet{jencson13ATEL} may have resolved the source at 5 GHz, finding major and minor axes of $2.6 \pm 0.6$ and $1.5 \pm 0.2$ mas, respectively. These correspond to physical sizes of $0.81 \pm 0.19$ and $0.47 \pm 0.06$ pc at NGC~2617.

\section{Analysis}
\label{sec:Anal}

We use our time-series spectra to search for variations in the broad H$\beta$ and \HeII$\lambda 4686$ lines and to estimate the central BH mass in \S\ref{sec:SpecAnal}. We cross-correlate our photometric light-curves in \S\ref{sec:timed}.  Finally, we construct a spectral energy distribution (SED) of NGC~2617 from the hard X-rays through the radio in \S\ref{sec:sed}.

\subsection{Emission-Line Variability and Black Hole Mass Estimate}
\label{sec:SpecAnal}

Accurate flux calibration is necessary to analyze intrinsic variations in AGN spectra. We therefore can only search for variability using the six well-calibrated, high signal-to-noise ratio (SNR) spectra that were collected under similar observing conditions (e.g., seeing, airmass; \citealp{peterson95}). These six spectra are marked with asterisks in Table~\ref{tab:spec} and Figure~\ref{fig:optical_spec}. We restricted our analysis to the \Hb{} emission-line region, since this spectral region is covered by all six spectra.  The region near \Ha{} is either outside the wavelength range or is located near the wavelength limits where the flux calibration is less accurate. Although the spectra in this subset were taken under similar observing conditions, there are still differences in the apertures and seeing that will lead to spectral differences. However, under the assumption that the \OIII{} narrow-line flux is constant over the time scale of interest (e.g., \citealp{peterson13}) and that the \OIII{} emitting region is not resolved, we can calibrate the spectra onto the same absolute scale using the observed \OIII$\lambda$5007 flux. The same is not true for the AGN continuum, which is contaminated by the emission of the host galaxy and, therefore, will still be subject to aperture effects. 

\begin{figure}
	\centerline{
		\includegraphics[width=17cm]{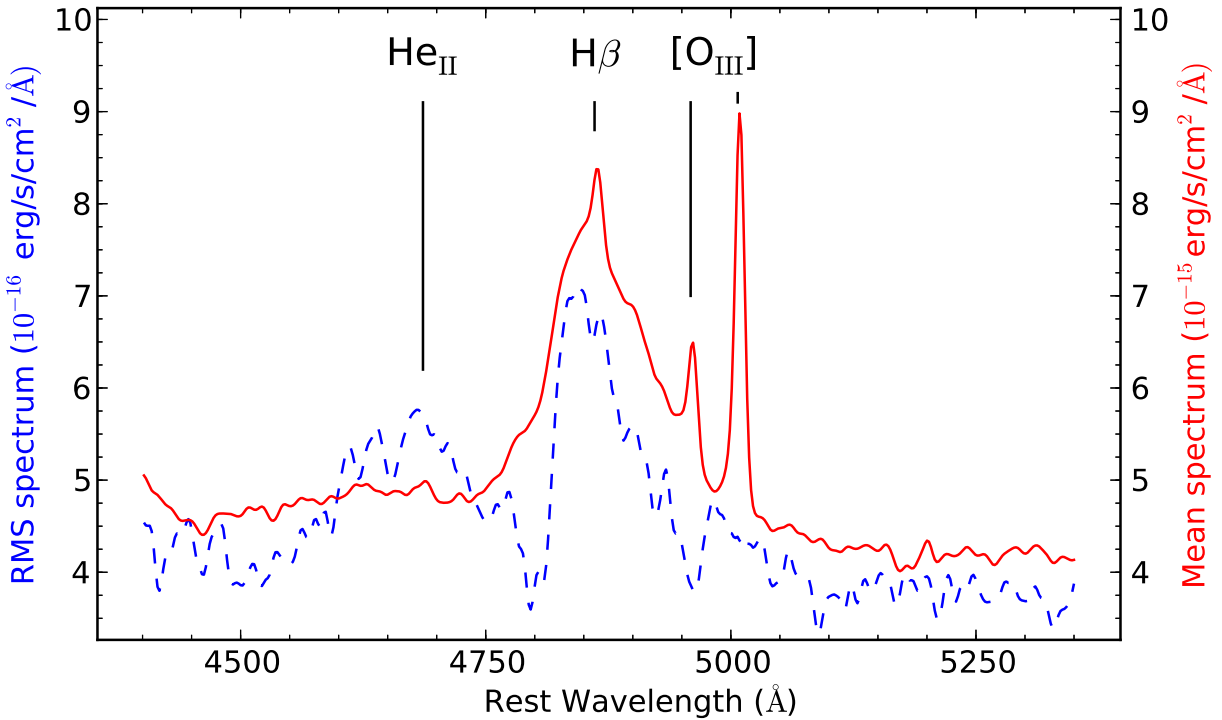}
}
	\caption{Mean (red solid, right flux scale) and RMS (blue dashed, left flux scale) spectra of NGC 2617 near the H$\beta$ and \HeII$\lambda 4686$ broad-lines.  Note that the RMS spectrum flux scale is 10 times lower than for the mean spectrum.  The RMS spectrum H$\beta$ and \HeII$\lambda 4686$ FWHM velocities are \HbFWHM{} and \HeIIFWHM{}, respectively, where the velocity dispersions are $\sigma = \Hbdisp{}$ and $\sigma = \HeIIdisp{}$, respectively. }
	\label{fig:spec_var}
\end{figure}

We degraded the six spectra to a common resolution of 9 \AA{} (FWHM). We adopted the spectrum taken on 2013  May 12.13 (under clear conditions) as the reference spectrum and inter-calibrated the spectra using the $\chi^2$ minimization algorithm of \citet{vangroningen92}, under the assumption that the \OIII$\lambda$5007 flux is constant. From the inter-calibrated spectra, we then  computed the mean and the root-mean-square (RMS) residual spectra shown in Figure~\ref{fig:spec_var}. The SNR of the RMS spectrum is very low due to the small number of spectra in the sample and the relatively small flux variations. Although a detailed analysis of the line profiles is precluded by the low SNR of the spectrum, both \Hb{} and \HeII{}  changed significantly during the period of interest. 

From the RMS spectrum, we can estimate the width of the variable part of the emission lines. Using a Monte Carlo approach, we generated 1000 sets of simulated spectra with randomized fluxes (see \citealp{peterson04}).  We then created an RMS spectrum for each simulated data set, and measured the FWHM and line dispersion ($\sigma_{\rm line}$) following \citet{peterson04}. Based on these procedures we find that the FWHM of the H$\beta$ and \HeII$\lambda 4686$ lines are \HbFWHM{} and \HeIIFWHM{} and their velocity dispersions are \Hbdisp{} and \HeIIdisp, respectively. The  uncertainties are relatively large due to the low SNR of the RMS spectrum. 

The width of the \Hb \ emission line and the mean continuum luminosity can then be used to estimate the mass of the black hole (\Mbh) powering the AGN. The width of the broad emission lines should primarily be due to the orbital motion of the emitting gas in the gravitational potential of the BH. Under this ``virial'' assumption, \Mbh \ can be estimated by 
\begin{equation}
M_{\rm{BH}}= f  \ G^{-1} \ R_{\rm BLR} \ \Delta v^2,
\end{equation}
where $\Delta v$ is the velocity dispersion of the emitting gas estimated from the width of the emission line, $R_{\rm BLR}$ is the distance of the broad line emitting gas from the central black hole, G is the gravitational constant, and $f$ is a dimensionless scale factor that depends on inclination, geometry and kinematics of the BLR (e.g., \citealp{onken04}).  Reverberation mapping studies of the \Hb \ emission line in local AGN indicate that there is a tight correlation between the AGN continuum luminosity at 5100 \AA \ and $R_{\rm BLR}$ \citep{kaspi05, bentz09, bentz13, zu11}.  Although we do not have enough spectral epochs to perform a reverberation mapping measurement, we can still estimate $R_{\rm BLR}$ through the radius-luminosity relation. Measuring $\lambda L_\lambda (5100$\AA$)$ from the mean spectrum and using the $R_{\rm BLR}-\lambda L_\lambda (5100$\AA$)$ relation from \citet{bentz13}, we obtain $R_{\rm BLR}=9\pm1$ light days.  Contamination by the host galaxy to the continuum luminosity can lead to an overestimate of the BLR radius.  However, we estimate in \S\ref{sec:Model} that the $V$-band contamination from the host galaxy in a 5$^{''}$ radius aperture is $\simeq 30\%$ of the flux. Since the spectroscopic apertures are smaller and $M_{\rm{BH}} \propto L^{1/2}$ our $R_{\rm BLR}$ estimate is little affected by host galaxy contamination.  Assuming $f=4.31\pm1.05$ \citep{grier13b} and using the \Hb \ $\sigma_{\rm line}$ estimate from the RMS spectrum, we obtain \Mbh{} $=$ \OurMbh.  For this estimate of the black hole mass, the Eddington luminosity is $L_{\rm{Edd}} = 10^{45.7}$ erg s$^{-1}$.

\subsection{Wavelength-Dependent Variability Lags and Temporal Smoothing}
\label{sec:timed}

\begin{figure*}
	\centerline{
		\includegraphics[height=16.5cm]{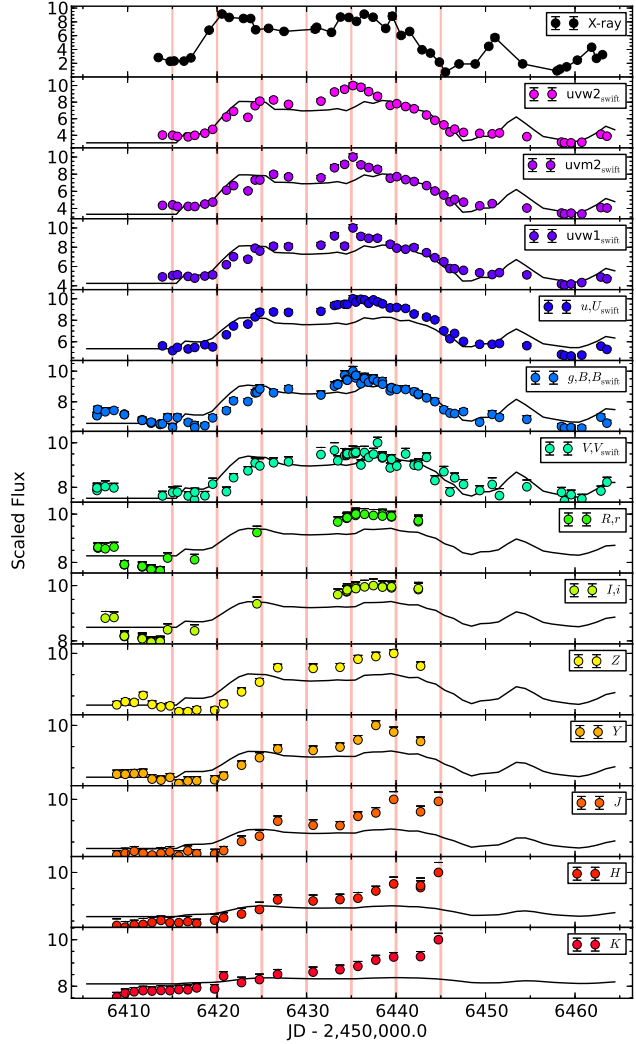}
	}
	\caption{Scaled NIR--UV photometric observations of NGC~2617 spanning $\sim60$ days from 2013 Apr. 24 through June 20. Similar filters have been scaled and combined into a common light curve.  Error bars are drawn but they are usually smaller than the points.  The coral vertical lines from 6415--6445 indicate the epochs where SEDs are shown in Figure~\ref{fig:sed} and aid in seeing the lag from shorter to longer wavelengths.  Black lines show light curves of a simple X-ray irradiation model described in \S\ref{sec:Model}.  While not a perfect match, the general amplitudes, lags, and smoothness of the light curves as a function of wavelength are successfully reproduced.  Finally, note how the relative strengths of the two peaks in the X-ray light curve change in the redward filters. This behavior is not captured in our our simple model with instantaneous reradiation. }
	\label{fig:flux}
\end{figure*}

Visual inspection of Figures~\ref{fig:phot} and \ref{fig:xray} clearly show a lag between the X-ray and UV--NIR variability.  It also seems reasonably clear that the longer-wavelength (e.g., NIR) light curves are also smoother than the shorter-wavelength light curves (e.g., UV).  In this subsection, we quantify these lags and smoothing time scales. A large number of our observations were obtained in filters that differ only slightly in wavelength. As a first step we combined the $u$ and $U_\textrm{swift}$ light curves; the $g$, $B$, and $B_\textrm{swift}$ light curves; the $V$ and $V_\textrm{swift}$ light curves; the $R$ and $r$ light curves; and the $I$ and $i$ light curves to improve the overall time sampling for each wavelength range.  To do this, we have added a constant to each band to match it to the partner band with the largest number of epochs.  Each of the resulting light curves are shown in Figure \ref{fig:flux} in arbitrary flux units and they are scaled to emphasize the light-curve structure. Figure \ref{fig:flux} further emphasizes the temporal ordering of the variability.

To quantitatively estimate the time lags and the temporal smoothing, we use the reverberation mapping and light curve analysis package JAVELIN\footnote{\url{https://bitbucket.org/nye17/javelin}}~(formerly known as SPEAR; \citealp{zu11, zu13, zu13b}). As with a normal cross-correlation analysis, JAVELIN assumes all emission line light curves are scaled and shifted versions of the continuum light curve. However, JAVELIN differs from simple cross-correlation methods in two respects.  First, JAVELIN explicitly builds a model of the light curve and transfer function and fits it to both the continuum and the line data, maximizing the likelihood of the model and then computing uncertainties using the (Bayesian) Markov Chain Monte Carlo~(MCMC) method. Second, as part of this process it models the continuum light curve using a damped random walk (DRW) model, which has been demonstrated to be a good statistical model of optical quasar variability using large ($\sim 10^4$) samples of quasar light curves (e.g., \citealp{kelly09, kozlowski10, macleod10, zu13}).  The parameters of the DRW model are included in the fits and their uncertainties, as is a simple top-hat model of the transfer function and the light curve means (or trends if desired).

Following \citet{zu11}, we apply the JAVELIN method to each of the light curves shown in Figure~\ref{fig:flux}, using either the \swift\ XRT X-ray or the \swift\ UVOT $UVW2$ band light curve as the reference (``continuum'') light curve and the other light curves at longer wavelengths are the responding (``line'') light curve. We fit fluxes rather than the magnitudes, which makes the results independent of any contamination by the host galaxy since the mean flux is subtracted from the light curves as part of the analysis.  Also, to avoid a spurious peaks at $\sim 15$ days, the spacing between the distinct two peaks visible in light curves, we forced JAVELIN to only return lags of $< 10$ days.  Finally, since X-ray variability typically has a different power spectrum than the UV/optical bands \citep{cameron12}, modeling the X-ray light curve as a DRW process may underestimate the uncertainties on the time lag of the paired light curves. However, the estimate of the time lag is determined mainly by the variability features manifestly present in the light curves, so the result should be insensitive to the statistics assumed for the underlying variability models.  

\begin{figure}
	\centerline{
		\includegraphics[width=14cm]{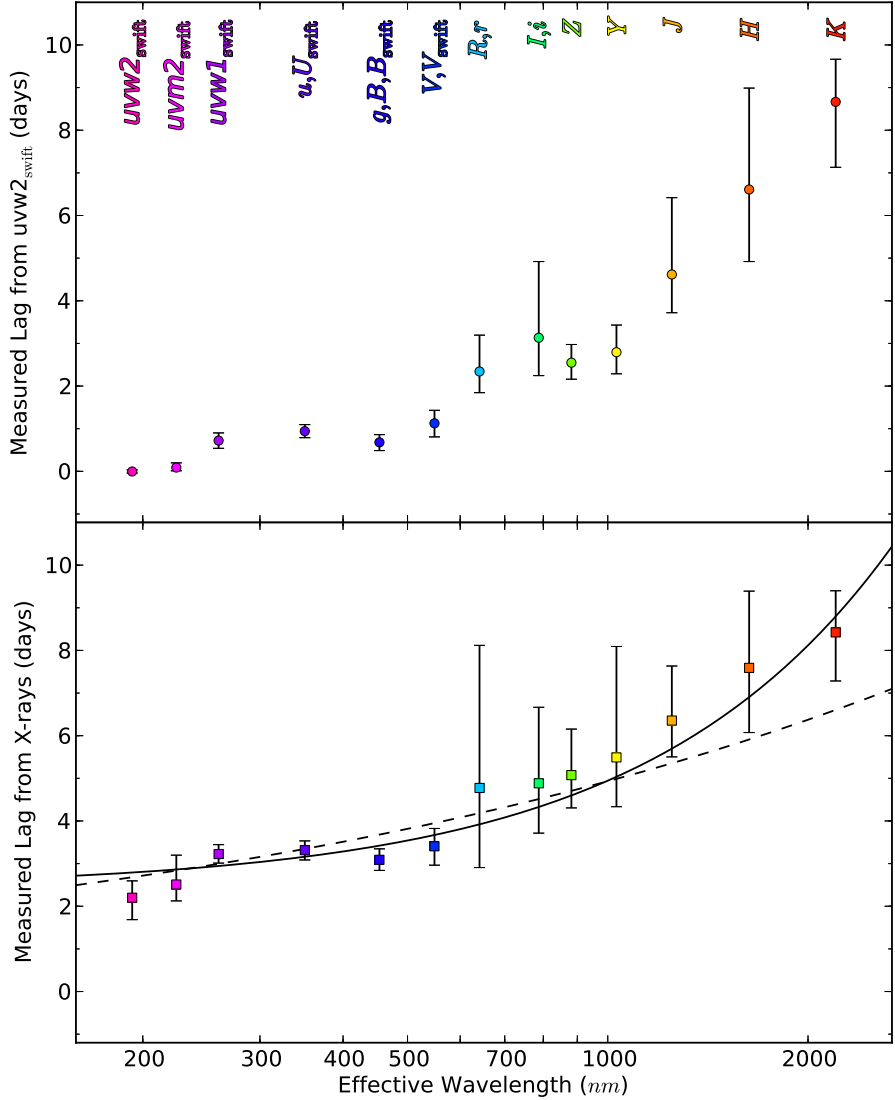}
}
	\caption{{\it Upper panel:} ({\it Lower panel:}) Measured time lags between variations in the \swift's UVOT $uvw2$-band (\swift's XRT X-ray) flux and the redder photometric bands as a function of wavelength. There is a clear trend that redder bands have larger lags, with the UV, optical, and NIR lagging $2\--3$, $3\--6$, and $6\--9$ days behind the X-rays, respectively.  The black dashed and solid lines show the best-fit power law and best-fit power law with a zero-point offset, respectively.}
	\label{fig:lag}
\end{figure}

The measured time lags are presented in Table~\ref{tab:lag}.  The upper panel (lower panel) of Figure~\ref{fig:lag} shows the measured time lags between the \swift's UVOT $uvw2$-band (\swift's XRT X-ray) flux and the redward photometric bands' flux as a function of wavelength.  There is a clear trend that redder bands have larger lags, with the UV and NIR lagging $2\--3$ and $6\--9$ days behind the X-rays, respectively.  We then fit a power law and a power law with a zero-point offset to the X-ray lags. The best-fits are
\begin{equation}
t_{\rm{lag}} = (\xlaga \pm \xlagaerr) \times \left( \frac{\lambda}{nm} \right)^{\xlagk \pm \xlagkerr}  \ \rm{days}
\end{equation}
and
\begin{equation}
t_{\rm{lag}} = \left( (\xlagbzero \pm \xlagbzeroerr) + (\xlagazero \pm \xlagazeroerr) \times \left( \frac{\lambda}{\mu m} \right)^{\xlagkzero \pm \xlagkzeroerr}\right)  \ \rm{days}
\label{eq:lagfitwithzero}
\end{equation}
where $t_{\rm{lag}}$ is the time lag as a function of wavelength ($\lambda$), respectively. The chi-square per degree of freedom for the fits are $\xlagredchisq$ and $\xlagredchisqzero$, respectively.  The $F$-test confidence level for the addition of the temporal offset parameter is $94.2\%$.  The zero-point of the second fit ($\xlagbzero \pm \xlagbzeroerr$ days) is in agreement with the non-geometric delay required by the simple quantitative model presented in \S\ref{sec:Model} to reproduce the observed UV-NIR light curves.  Furthermore, depending on whether dissipation or irradiation dominates, we would expect $t_{\rm{lag}} \propto \lambda^\alpha$ with $4/3 \leq \alpha \leq 2$ (see Equation \ref{eq:toymodel}).  The results for Equation \ref{eq:lagfitwithzero} are consistent with the shallower slopes where dissipation remains the dominant heating mechanism.

\input{tablag.tex}

\begin{figure}
	\centerline{
		\includegraphics[width=14cm]{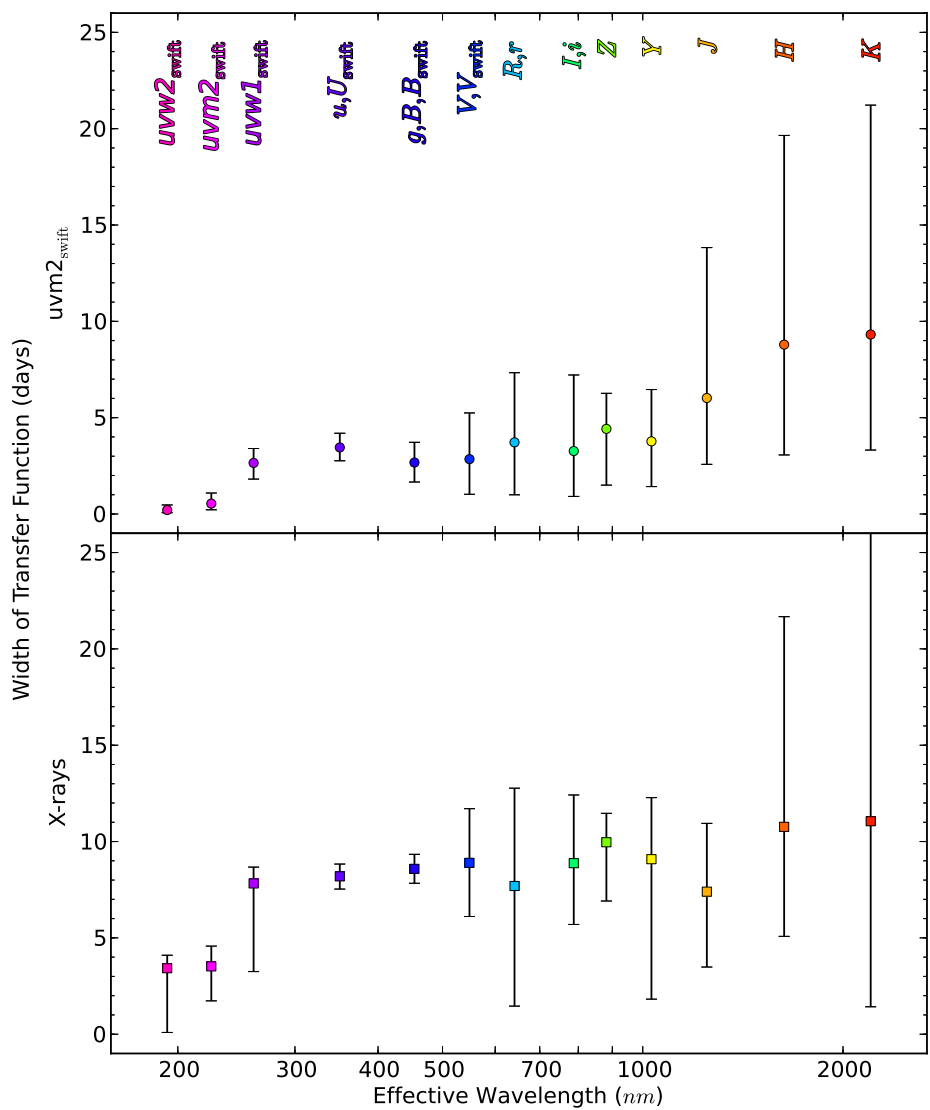}
}
	\caption{{\it Upper panel:} ({\it Lower panel:}) Measured widths for the transfer function between the \swift's UVOT $uvw2$-band (\swift's XRT X-ray) flux and the redder photometric bands as a function of wavelength. There is a reasonably clear trend that redder bands have wider transfer functions and thus smoother light curves.}
	\label{fig:smooth}
\end{figure}

The JAVELIN models also smooth the reference light curves with a top hat filter, and the wavelength-dependent top hat widths are presented in  Table~\ref{tab:lag} and shown in Figure~\ref{fig:smooth}.   The upper panel (lower panel) of Figure~\ref{fig:smooth} shows the width of the transfer function between the \swift's UVOT $uvw2$-band (\swift's XRT X-ray) flux and the redward photometric bands' flux as a function of wavelength.  Even though the widths become increasingly uncertain at longer wavelengths, there is still a clear trend that redder bands have larger widths, and thus smoother light curves. If the X-ray to UV lag and transfer function width measurements are robust, they have interesting implications for the origin of variability in AGN which are discussed in \S\ref{sec:Model}.

\subsection{Spectral Energy Distribution}
\label{sec:sed}

\begin{figure}
	\centerline{
		\includegraphics[width=9cm]{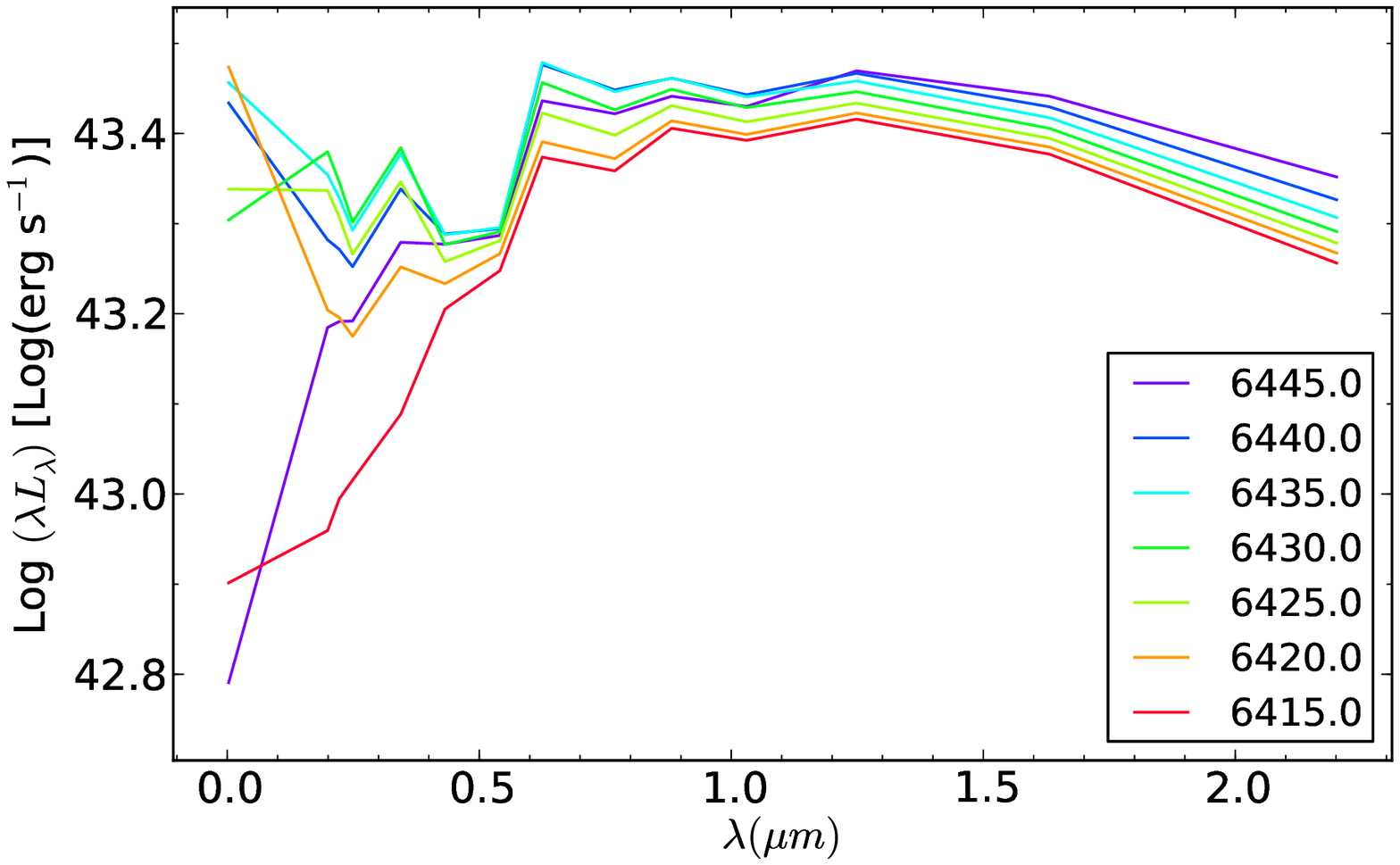}
		\includegraphics[width=9cm]{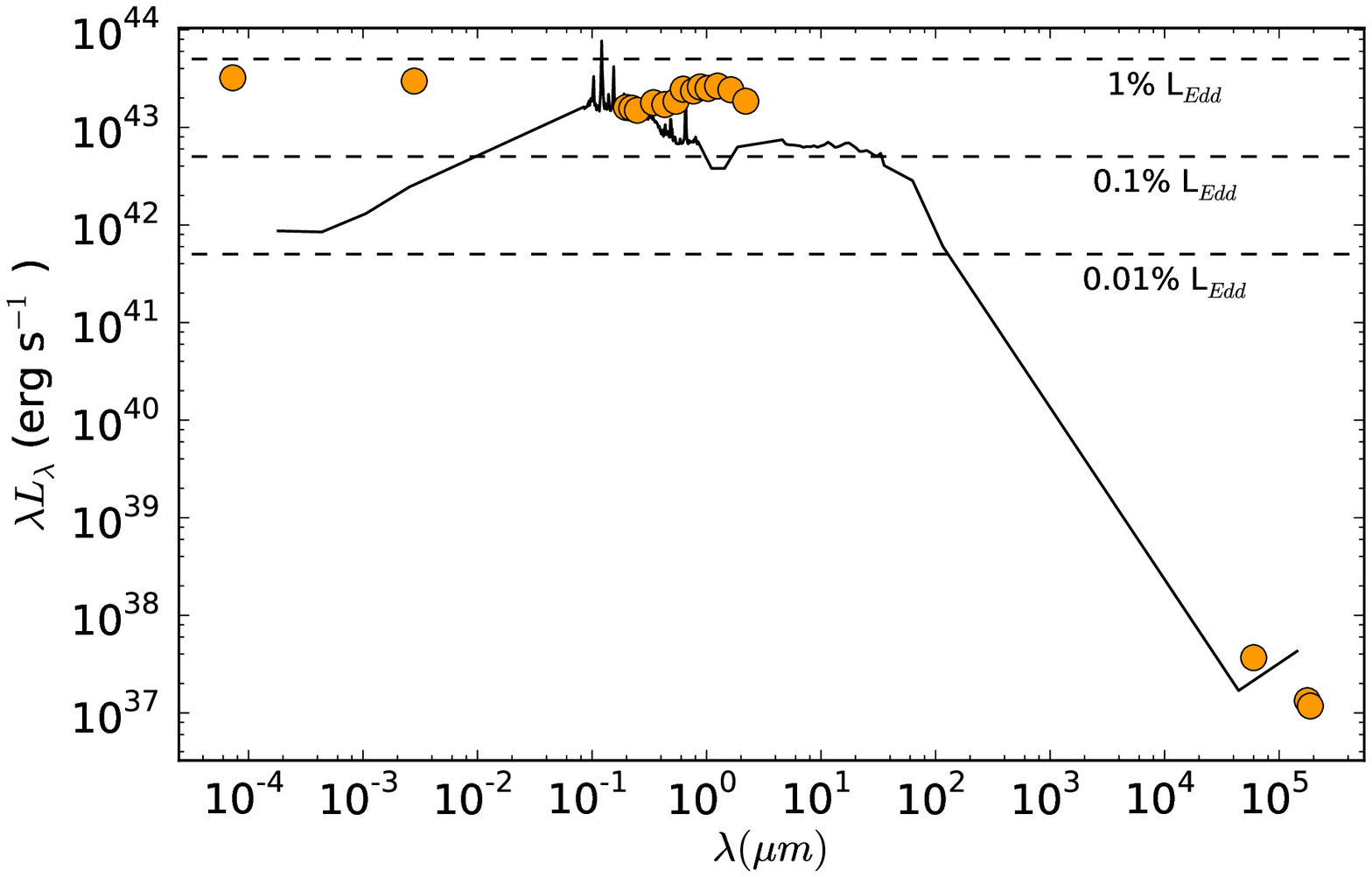}
}
	\caption{SED of NGC 2617.  {\it Left panel:} SED as a function of time as determined from our combined light curves shown in Figure~\ref{fig:flux}.  {\it Right panel:} SED stretching from hard X-rays to Radio. X-ray through NIR observations were interpolated on the JD date 2,456,420.0.  The Integral $17\--60$~keV X-ray observation was obtained during this epoch.     Radio observations shown were obtained 30--52 days later. The black line shows the composite SED for radio-quiet quasars from \citet{shang11} scaled to match in the UV. Fractions of $L_{\rm{Edd}}$ assuming $\log M_{\rm{BH}}/M_\odot = 7.6$ are shown for reference. The left panel shows that the X-ray and UV fluxes rapidly strengthen at the beginning of the flare. The right panel shows that NGC 2617 was extremely X-ray bright during the flare, NGC 2617 is radio quiet, and our SED contains a significant amount of contamination from the host galaxy in the optical and NIR wavelength ranges (see Figure~\ref{fig:Modelsed}). }
	\label{fig:sed}
\end{figure}

Using the combined fluxes from \S\ref{sec:timed}, the Integral $17\--60$~keV observation discussed in \S\ref{sec:HardX}, and the EVN and VLBA radio observations discussed in \S\ref{sec:Radio} we produce SEDs of NGC~2617. To put the X-ray observations on this SED we assigned an effective wavelength for the \swift\ X-ray band, using the measured energy spectral index, and for the Integral X-ray band, assuming a flat energy spectral index.  We then multiplied the measured X-ray fluxes by this effective wavelength and divided by the wavelength range integrated in each measurement.  Figure~\ref{fig:sed} shows the X-ray through NIR SED of NGC 2617 at a number of epochs in the left panel and a hard X-ray through radio SED in the right panel.  For the right panel of Figure~\ref{fig:sed}, we interpolated the X-ray through NIR light curves on the JD$=2,456,420.0$, we show the Integral $17\--60$~keV X-ray observation which was obtained during this epoch, and we show the EVN and VLBA observations taken 30 and 52 days later, respectively. During the beginning of the flare there was a marked increase in the X-ray and UV flux while the increase at longer wavelengths is less pronounced.  

In the right panel of Figure~\ref{fig:sed} we also show the composite SED for radio-quiet quasars of \citet{shang11} scale to match in the \swift\ UV fluxes.  We normalized the SED to the UV luminosity to minimize the effects of host-galaxy contamination which, given the shape of the SED in the optical--NIR, is large. In \S\ref{sec:Model} we include a model of this contamination and find it is roughly consistent with the SED of an early type spiral. Finally, we see that NGC~2617 was extremely X-ray bright during the flare, being 10--25 times brighter than the composite SED. 

In Figure~\ref{fig:sed}, we also see that the SED roughly has $\nu L_{\nu} \sim 0.01 L_{\rm{Edd}}$, although there is significant host contamination as we move to the redder optical and NIR bands (see \S\ref{sec:Model} and Figure~\ref{fig:Modelsed}).  Assuming the existence of an unobserved hard UV accretion peak, this suggests that the AGN flare peaked at $\sim 10\%$ of $L_{\rm{Edd}}$.  This estimate is consistent estimates ($\sim$0.03--0.15) obtained using Equation 1 from \citet{grupe10} and the measured X-ray spectral indices reported in Table~\ref{tab:phot} and shown in Figure~\ref{fig:xray}.

\section{A Simple Quantitative Model}
\label{sec:Model}

Since the sequence of variability seems to lead with the X-rays, followed by the ultraviolet and eventually the NIR, the natural explanation of the observations seems to be variability driven by X-ray irradiation of the BH accretion disk.  We know from gravitational microlensing of lensed X-ray emission from quasars (e.g., \citealt{mosquera13}) that the X-ray emission has a half-light radius of order ten times the gravitational radius ($r_g$).  This is small enough, $10 r_g/c \simeq 0.5$~hours, such that we can treat the X-ray emission as a point source. Therefore, we consider a simple ``lamp post'' model (e.g., \citealp{berkley00}, \citealp{kazanas01}) for the NGC~2617 outburst which consists of a flat, non-relativistic thin disk with an accretion luminosity $L_A$ and an inner disk edge at $R_{in}=3r_g$ illuminated by an X-ray point source a distance $h_X$ above the disk center. 

In this model we assume that all X-rays incident on the disk are absorbed and instantaneously heat the disk surface.  In this case the local disk temperature at observed time $t$ is 
\begin{equation}
   \sigma T(R,t)^4 = { \frac{3 L_A R_{in}}{2 \pi  R^3}} \left[ 1- \left( { \frac{R_{in}}{ R }}\right)^{1/2}\right]
        + { \frac{L_X(t-\tau(R))}{4 \pi \left( R^2 + h_X^2 \right)} } { \frac{h_X}{\left(R^2 + h_X^2\right)^{1/2}} } 
\label{eq:toymodel}
\end{equation}
where $L_X(t)$ is the X-ray luminosity and $\tau(R) = \tau_0 + h_X/c + \left(R^2+h_X^2\right)^{1/2}/c$ is (for simplicity) the time delay between light from radius $R$ and the direct X-ray emission. To synthesize emission at wavelength $\lambda$ we simply assume black body emission at each radius and integrate over the disk,
\begin{equation}
   \nu F_\nu = { \frac{4 \pi h \nu^4}{c^2 D^2} } \int_{R_{in}}^\infty { \frac{R dR}{\exp(h\nu/kT(R,t))-1} }.
\end{equation}  
We experimented with adding a disk opening angle and a bolometric correction for the X-ray fluxes, but these had little effect.  

As seen in \S\ref{sec:sed}, our fairly large aperture photometry contains significant host contamination.  We experimented with several approaches to remove this contamination short of fully modeling the images.  For example, the contamination is reasonably well modeled by the Sbc SED template from \citet{assef10}.   However, there are systematic offsets that still dominate the $\chi^2$ of the resulting fits, which could either be due to the host or our use of a simple, non-relativistic thin disk model.  Since such questions are peripheral to the temporal changes in the fluxes, we simply included an independent, constant ``host luminosity'' for each band found by doing a $\chi^2$ fit for fixed AGN parameters. We then experimented with the parameters for the accretion disk luminosity $L_A$, the ``quiescent'' X-ray luminosity $L_{X0}$ used where we lack Swift data, the X-ray source height $h_X$ and the non-geometric delay $\tau_0$ keeping the black hole mass fixed at $\log{(M_{\rm{BH}}/M_\odot)} = 7.6$.

\begin{figure}
\centerline{\includegraphics[width=5.5in]{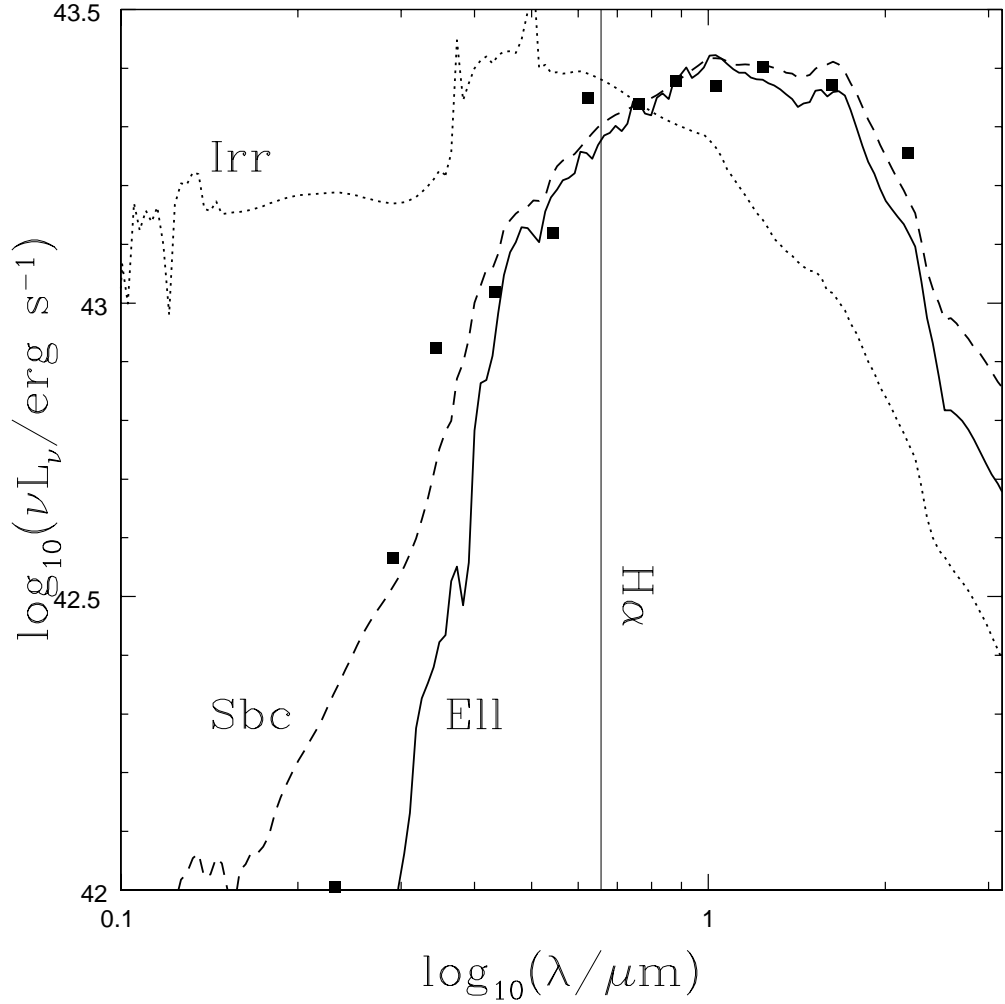}}
\caption{
  The SED for the ``host galaxy'' contamination in the best fit model as compared to the elliptical (solid),   Sbc (dashed) and irregular (dotted) galaxy templates from \citet{assef10}. The host is well matched to an  SED to the Sbc template, with deviations near H$\alpha$.
  }
\label{fig:Modelsed}
\end{figure}

Given the simplicity of the model, we can obtain surprisingly good (but not perfect) fits.  First, Figure~\ref{fig:Modelsed} shows the SED of the ``host galaxy'' contamination model as compared to the SED templates of \citet{assef10}.  Although we did not impose any prior that these estimates resemble the SED of a galaxy, we see that the estimates are broadly consistent with the Sbc template except near the H$\alpha$ line.  Because the broad lines should respond less and more slowly than the continuum, it is not surprising that they are absorbed into the host-galaxy parameters.  The match to the galaxy templates (or combinations of them) is not perfect due to both observational issues (e.g. seeing variations between observations will lead to variations in the aperture flux of the host) and any shortcomings in our very simple disk model.  There is also a tendency to over-estimate the host flux at the longest wavelengths. 

Figure~\ref{fig:flux} shows the model UV, optical and NIR light curves compared with the observed light curves. The model naturally reproduces the patterns seen in the data, namely that the shorter wavelengths have shorter lags and are less smooth than longer wavelengths, simply because of geometry. Shorter wavelength emission comes from smaller radii with shorter light crossing times to the X-ray source, leading to the correlation of the lags with wavelength.  Furthermore, the size of the region emitting the shorter wavelengths is smaller, so there is less temporal smoothing of the X-ray emission.  This is somewhat exaggerated at the longer wavelengths because the host galaxy contribution also reduces the fractional variability, which is why the analysis in \S\ref{sec:timed} was carried out using fluxes rather than magnitudes. For example, the tendency to overshoot the NIR host flux leads to the less than observed NIR variability seen in Figure~\ref{fig:flux}.

The best fit model has $\log{L_A}=42.6$ and $\log{L_{X0}}=42.6$ (in ergs~s$^{-1}$) which is $10^{-3.1} L_{Edd}$ for the estimated black hole mass.  The X-ray source can be at a very low altitude, $\log(h_x/\hbox{cm})=13.5$ which is $4.8 r_g$ where $r_g=GM_{BH}/c^2$ is the gravitational radius of the black hole, consistent with the results from microlensing (e.g. \citealt{mosquera13}). We do not think an attempt to assign uncertainties to the model parameters is presently warranted.  

The three main weaknesses of our model's fit to the data are:  
1) The model NIR light curves provide a poor fit to the observed data. This could be due to a number of reasons.  First, as mentioned above, accounting for the host galaxy flux is especially important in the NIR where the galaxy contribution is large and our over estimation of the host galaxy's contribution at the longest wavelength reduces the expected variability.  Second, our model does not include any dust reprocessing. If there is significant reprocessing of UV and optical photons on scales larger than the BLR, then the NIR would respond to the increase in X-ray flux on two distinct timescales, whereas only one is currently included in our models. 
2) We do not reproduce the second bump seen in the X-ray light curve around JD=2,456,451 in the UV--NIR light curves. The reason for this is unclear. 
3) We were unable to find successful solutions without the non-geometric lag $\tau_0 \simeq 2$~days.   Trying to make it part of the geometric lag by putting the X-ray source at $h_x \simeq \tau_0/c$ above the disk did not work and would be an odd geometry in any case, with the X-ray emission arising from $\sim 10^3 r_g$ above the disk.  The basic problem is that the UV emission arises from such small radii that there is no geometric lag relative to a central X-ray source, so that portion of the delayed response has to be supplied by $\tau_0$.  
We do not have a physical interpretation for this non-geometric lag.

\section{Summary}
\label{sec:SumTotal}

On 2013  Apr. 10.27, ASAS-SN was triggered  by a $\sim10$\% relative flux increase from the inner regions of NGC~2617. Follow-up spectroscopy found that NGC~2617, previously classified as a Seyfert 1.8 galaxy, had also changed its spectral type to become a Seyfert 1 with broad optical emission lines and a continuum blue bump.  We then began a $\sim70$ day photometric and spectroscopic monitoring campaign from the NIR -- X-rays, observing NGC~2617 through a dramatic outburst in which its X-ray flux increased by over an order of magnitude followed by an increase in its optical/ultraviolet (UV) continuum flux by almost an order of magnitude.  Although these spectral changes could have occurred at any time in the intervening decade, it seems most likely that the change in Seyfert type was associated with the strong outburst given the rarity of both ``changing-look'' AGN and powerful X-ray flares.


When analyzing these observations we found significant variations in the broad emission lines H$\beta$ and \HeII$\lambda 4686$.  When cross-correlating our X-ray light curve and the other photometric light-curves we found that the disk emission lags the X-rays, with the lag becoming longer as we move from the UV ($2\--3$ days) to the NIR ($6\--9$ days).  We also see that the NIR is more heavily temporally smoothed than the UV. Since the sequence of variability seems to lead with the X-rays and is followed by redward bands, the natural explanation of the observations seems to be variability driven by X-ray irradiation of the disk, possibly due to coronal activity.  A simple physical model using X-ray illumination to heat a thin disk roughly reproduces the UV--NIR light curves.  That such a simple model works at all strongly suggests that a more detailed and realistic modeling study is warranted. 

We believe this is the first case where there is a clear and convincing determination of whether X-ray variability drives UV$\rightarrow$NIR variability or the reverse.  In this outburst, the X-ray variability unambiguously drives the UV$\rightarrow$NIR variability, almost certainly by irradiating and heating the disk.  While quasar and AGN variability is ubiquitous, it has proved difficult to measure time delays between thermal disk and non-thermal X-ray emission or between different wavelengths of the thermal emission, leaving the origin of the variability unclear.  We can now quantify quasar variability well, using the DRW model (e.g. \citealt{kelly09}, \citealt{kozlowski10}, \citealt{macleod10}, \citealt{macleod12}, \citealt{zu13}) and the parameters of the DRW model are correlated with properties of the AGN.  Determining the physical origins of these parameters and their scalings has been more difficult. For example, \citet{kelly09} argue that the DRW time scale could be related to the thermal time scale of the disk, while \citet{macleod10} note that the scalings of both the DRW time scale and amplitude with wavelength, black hole mass, and luminosity all seem too weak compared to the natural scalings one would expect for a simple disk model.  A program systematically triggering high cadence \swift\ monitoring campaigns after optically detecting a significant luminosity change can likely provide similar results to this case for larger numbers of systems to test whether the pattern we saw here is unique or ubiquitous, and the time ordering of the variability should lead to a clear physical model for understanding quasar variability more generally.  Ideally, one should also try to have far better spectroscopic sampling than we achieved here so that one simultaneously carries out a reverberation mapping campaign to directly determine the size of the broad line region and black hole mass.

\acknowledgments

We thank the APO director, S. Hawley, for granting us DD Time for this observation, and the observatory crew.  Additionally, we would like to thank N. Gehrels for approving our \swift\ ToO  observation requests and the \swift\ ToO team for promptly scheduling and executing these observations.  We also thank Jennifer van Saders, Todd Thompson, Michael Fausnaugh, and Joe Antognini for discussions and encouragement. We also thank the anonymous referee for his/her helpful comments and suggestions improving this manuscript.  B.J.S. was supported by a Graduate Research Fellowship from the National Science Foundation. Development of ASAS-SN has been supported by NSF grant AST-0908816 and the Center for Cosmology and AstroParticle Physics at The Ohio State University.  \swift\ at PSU is supported by NASA contract NAS5-00136.  C.S.K. and B.J.S are supported by NSF grant AST-1009756. G.D.R, B.M.P, and R.W.P. are supported by NSF grant AST-1008882.  S.K. would like to thank the Kavli Institute for Theoretical Physics (KITP) for their hospitality and support during the program on ``A Universe of Black Holes''. S.K. is supported by NSF grant NSF PHY11-25915.  M.I., C.C., and Y.Y. acknowledge the support from the Creative Initiative program No. 2008-0060544 of the National Research Foundation of Korea (NRFK) funded by the Korea government (MSIP). J.F.B. is supported by NSF grant PHY-1101216. This paper uses the data taken by the United Kingdom Infrared Telescope which is operated by the Joint Astronomy Centre on behalf of the Science and Technology Facilities Council of the U.K.  This paper includes data gathered with the 6.5 meter Magellan Telescopes located at Las Campanas Observatory, Chile. The SDSS is managed by the Astrophysical Research Consortium for the Participating Institutions. Funding for the SDSS and SDSS-II has been provided by the Alfred P. Sloan Foundation, the Participating Institutions, the National Science Foundation, the U.S. Department of Energy, the National Aeronautics and Space Administration, the Japanese Monbukagakusho, the Max Planck Society, and the Higher Education Funding Council for England. The SDSS Web Site is http://www.sdss.org/. This research has made use of the NASA/IPAC Extragalactic Database (NED) which is operated by the Jet Propulsion Laboratory, California Institute of Technology, under contract with the National Aeronautics and Space Administration. This research has made use of NASA's Astrophysics Data System Bibliographic Services. This paper makes use of data from the AAVSO Photometric All Sky Survey, whose funding has been provided by the Robert Martin Ayers Sciences Fund.

\bibliography{/home/petrus/shappee/Latex/references}
\bibliographystyle{mn}

\end{document}

%% file: xlag_fit.tex
\newcommand{\xlaga}{\ensuremath{0.38}} 
\newcommand{\xlagaerr}{\ensuremath{0.19}} 
\newcommand{\xlagk}{\ensuremath{0.37}} 
\newcommand{\xlagkerr}{\ensuremath{0.08}} 
\newcommand{\xlagredchisq}{\ensuremath{1.48}} 
\newcommand{\xlagazero}{\ensuremath{2.5}} 
\newcommand{\xlagazeroerr}{\ensuremath{0.7}} 
\newcommand{\xlagkzero}{\ensuremath{1.18}} 
\newcommand{\xlagkzeroerr}{\ensuremath{0.33}} 
\newcommand{\xlagbzero}{\ensuremath{2.43}} 
\newcommand{\xlagbzeroerr}{\ensuremath{0.46}} 
\newcommand{\xlagredchisqzero}{\ensuremath{0.97}}

%% file: tabphot.tex
\begin{deluxetable}{lrcc}
\tablewidth{350pt}
\tabletypesize{\footnotesize}
\tablecaption{Photometric Observations}
\tablehead{
\colhead{JD} &
\colhead{} &
\colhead{Magnitude | X-Ray Flux} &
\colhead{Photon Index} \\ 
\colhead{($-$2,450,000)} &
\colhead{Band} &
\colhead{ \ \ \ \ \ \ \ \ \ \ \ \ \ \ \ \ \ \ \    \nodata | [$10^{-11}$ ergs/cm$^{2}$/s]}  &
\colhead{$\Gamma$}  }
\startdata
6413.430 & X-Ray$_\mathrm{swift}$ & [2.84(0.12)] & 1.69(0.10) \\ 
6413.896 & uvw2$_\mathrm{swift}$ & 14.48(0.03) & \nodata \\ 
6413.900 & uvm2$_\mathrm{swift}$ & 14.32(0.04) & \nodata \\ 
6413.891 & uvw1$_\mathrm{swift}$ & 14.17(0.04) & \nodata \\ 
6413.892 & $U_\mathrm{swift}$ & 14.22(0.03) & \nodata \\ 
6433.483 & $u$ & 14.69(0.03) & \nodata \\ 
6406.570 & $B$ & 14.92(0.03) & \nodata \\ 
6413.893 & $B_\mathrm{swift}$ & 15.08(0.03) & \nodata \\ 
6406.645 & $g$ & 14.55(0.03) & \nodata \\ 
6406.575 & $V$ & 14.32(0.03) & \nodata \\ 
6413.898 & $V_\mathrm{swift}$ & 14.54(0.03) & \nodata \\ 
6406.683 & $r$ & 13.88(0.02) & \nodata \\ 
6434.472 & $R$ & 13.58(0.02) & \nodata \\ 
6407.499 & $i$ & 13.69(0.03) & \nodata \\ 
6434.475 & $I$ & 13.13(0.02) & \nodata \\ 
6408.787 & $Z$ & 12.96(0.01) & \nodata \\ 
6408.785 & $Y$ & 12.71(0.01) & \nodata \\ 
6408.782 & $J$ & 12.18(0.01) & \nodata \\ 
6408.780 & $H$ & 11.51(0.03) & \nodata \\ 
6408.777 & $K$ & 11.05(0.03) & \nodata \\ 
\enddata \tablecomments{\swift, $U$, $B$, $V$, $R$, $I$, $Z$, $Y$, $J$, $H$, and $K\--\textrm{band}$ photometry are calibrated in the Vega magnitude system. $u$, $g$, $r$, and $i\--\textrm{band}$ photometry are calibrated in the AB magnitude system.  X-Ray fluxes (0.3$\--$10 keV) are in units of $10^{-11}$ ergs/cm$^{2}$/s. \textit{Only the first observation in each band is shown here to demonstrate its form and content. Table to be published in its entirety in machine-readable form.}} 
\label{tab:phot} 
\end{deluxetable}

%% file: tabspec.tex
\begin{deluxetable}{lclcrccccr}
\tabletypesize{\tiny}
\tablewidth{450pt}
\tablecaption{Spectroscopic Observations}
\tablehead{\colhead{} & 
\colhead{JD} &
\colhead{} &
\colhead{Wavelength range} &
\colhead{Resolution} &
\colhead{Slit Width} &
\colhead{Seeing} &
\colhead{Position Angle} &
\colhead{} &
\colhead{Exposure} \\
\colhead{UT Date} &
\colhead{-2,400,000} &
\colhead{Telescope/Instrument} &
\colhead{(\AA)} &
\colhead{(\AA)} &
\multicolumn{2}{c}{arcsec} &
\multicolumn{1}{c}{degrees} &
\colhead{Airmass} &
\colhead{(s)}   }
\startdata

2013 Apr 25.10* & 56407.601 &  APO3.5m/DIS            &  $3500\--9600$&           9.0    &      1.5  & 1.3 & 30  & 1.3 &    $2\times 300$ \\
2013 Apr 29.19* & 56411.693 &  APO3.5m/DIS*           &  $3500\--9600$&           9.0    &      1.5  & 1.5 & 50  & 2.1 &    $2\times 300$ \\
2013 Apr 30.30 & 56412.800 &  UH2.2m/SNIFS           &  $3300\--9700$&           5.0    &   \nodata      & $\sim0.9$ & \nodata & 1.5 &  $1\times 600$ \\
2013 May 12.13* & 56424.629 &  MDM2.4m/CCDS          &  $4250\--5850$&           4.0    &      1.0  & 1.7 & 0  & 1.5 &    $2\times 600$ \\
2013 May 19.15* & 56431.647 &  MDM2.4m/OSMOS          &  $3970\--6850$&           4.4    &      1.2  & 1.7 & 0  & 1.9 &    $2\times 600$ \\
2013 May 20.14* & 56432.643 &  MDM2.4m/OSMOS          &  $3970\--6850$&           4.9    &      1.2  & 2.6 & 0  & 1.9 &    $2\times 600$ \\
2013 May 21.14* & 56433.637 &  MDM2.4m/OSMOS          &  $3970\--6850$&           5.0    &      1.2  & 1.8 & 0  & 1.9 &    $2\times 600$ \\
2013 May 27.41 & 56439.912 &  FTS/FLOYDS          &  $3200\-- 10000$&            9.7   &     2.0  & $\sim1.0$ & 125.24 &  1.8 &    $1\times 1800$ \\  
2013 Jun 1.14 & 56444.644 &  MDM1.3m/CCDS           &  $4200\--6200$&           14.4   &      1.9   & 5.6 & 0  & 2.7 &  $2\times 900$ \\
2013 Jun 2.14 & 56445.641 &  MDM1.3m/CCDS           &  $4200\--6200$&           14.8   &      1.9   & 7.7 & 0  & 2.7 &  $1\times 900$ \\
2013 Jun 3.14 & 56446.645 &  MDM1.3m/CCDS           &  $4200\--6200$&           14.5   &      1.9   & 6.7 & 0  & 2.9 &  $2\times 900$ \\
2013 Jun 4.14 & 56447.642 &  MDM1.3m/CCDS           &  $4200\--6200$&           13.8   &      1.9   & 5.6  & 0  & 2.9 &  $2\times 900$ \\
2013 Jun 30.96 & 56474.456 &  Magellan~6.5m/IMACS    &  $4000\--9600$&            5.2   &      0.9  & 1.7  & 117  & 2.4 &  $1\times 800$ \\

\enddata 
\tablecomments{Observational and derived properties of our time series spectra of NGC 2617. Seeing gives the FWHM of the spatial profile in the 2d spectra. The airmass is reported for the beginning of the exposures.  Spectra with UT Date marked with a * are used in the spectral analysis presented in \S \ref{sec:SpecAnal}.}

\label{tab:spec}

\end{deluxetable}

%% file: tablag.tex
\begin{deluxetable}{lcccc}
\tablewidth{200pt}
\tabletypesize{\small}
\tablecaption{Measured Time Lags and Smoothing Scales}
\tablehead{
\colhead{} &
\colhead{UV lag} &
\colhead{X-Ray lag} &
\colhead{UV smoothing} &
\colhead{X-Ray smoothing} \\ 
\colhead{Band} &
\multicolumn{2}{c}{days}  &
\multicolumn{2}{c}{days}  } 
\startdata
$uvw2_\mathrm{swift}$ & $ -0.00^{+ 0.04 }_{- 0.04 } $ & $ 2.20^{+ 0.51 }_{- 0.39 } $ & $ 0.21^{+ 0.14 }_{- 0.26 } $ & $ 3.42^{+ 3.33 }_{- 0.67 }  $ \\ 
$uvm2_\mathrm{swift}$ & $ 0.09^{+ 0.08 }_{- 0.11 } $ & $ 2.51^{+ 0.38 }_{- 0.69 } $ & $ 0.54^{+ 0.32 }_{- 0.55 } $ & $ 3.52^{+ 1.79 }_{- 1.05 }  $ \\ 
$uvw1_\mathrm{swift}$ & $ 0.72^{+ 0.18 }_{- 0.18 } $ & $ 3.22^{+ 0.21 }_{- 0.22 } $ & $ 2.65^{+ 0.84 }_{- 0.75 } $ & $ 7.82^{+ 4.57 }_{- 0.85 }  $ \\ 
$u,U_\mathrm{swift}$ & $ 0.94^{+ 0.15 }_{- 0.15 } $ & $ 3.32^{+ 0.23 }_{- 0.22 } $ & $ 3.45^{+ 0.69 }_{- 0.74 } $ & $ 8.19^{+ 0.66 }_{- 0.64 }  $ \\ 
$g,B,B_\mathrm{swift}$ & $ 0.68^{+ 0.19 }_{- 0.18 } $ & $ 3.09^{+ 0.25 }_{- 0.26 } $ & $ 2.67^{+ 1.01 }_{- 1.05 } $ & $ 8.58^{+ 0.75 }_{- 0.75 }  $ \\ 
$V,V_\mathrm{swift}$ & $ 1.13^{+ 0.32 }_{- 0.31 } $ & $ 3.41^{+ 0.44 }_{- 0.42 } $ & $ 2.85^{+ 1.82 }_{- 2.40 } $ & $ 8.90^{+ 2.79 }_{- 2.81 }  $ \\ 
$R,r$ & $ 2.34^{+ 0.50 }_{- 0.85 } $ & $ 4.78^{+ 1.87 }_{- 3.34 } $ & $ 3.72^{+ 2.72 }_{- 3.62 } $ & $ 7.69^{+ 6.23 }_{- 5.08 }  $ \\ 
$I,i$ & $ 3.13^{+ 0.89 }_{- 1.79 } $ & $ 4.88^{+ 1.17 }_{- 1.78 } $ & $ 3.27^{+ 2.35 }_{- 3.95 } $ & $ 8.88^{+ 3.18 }_{- 3.54 }  $ \\ 
$Z$ & $ 2.55^{+ 0.39 }_{- 0.43 } $ & $ 5.08^{+ 0.77 }_{- 1.08 } $ & $ 4.42^{+ 2.92 }_{- 1.85 } $ & $ 9.96^{+ 3.05 }_{- 1.50 }  $ \\ 
$Y$ & $ 2.79^{+ 0.51 }_{- 0.64 } $ & $ 5.49^{+ 1.16 }_{- 2.60 } $ & $ 3.77^{+ 2.35 }_{- 2.69 } $ & $ 9.08^{+ 7.26 }_{- 3.20 }  $ \\ 
$J$ & $ 4.62^{+ 0.90 }_{- 1.80 } $ & $ 6.36^{+ 0.85 }_{- 1.28 } $ & $ 6.02^{+ 3.44 }_{- 7.81 } $ & $ 7.40^{+ 3.91 }_{- 3.54 }  $ \\ 
$H$ & $ 6.61^{+ 1.69 }_{- 2.38 } $ & $ 7.59^{+ 1.52 }_{- 1.80 } $ & $ 8.79^{+ 5.73 }_{- 10.86 } $ & $ 10.76^{+ 5.68 }_{- 10.91 }  $ \\ 
$K$ & $ 8.66^{+ 1.53 }_{- 1.00 } $ & $ 8.42^{+ 1.14 }_{- 0.98 } $ & $ 9.32^{+ 6.00 }_{- 11.91 } $ & $ 11.05^{+ 9.63 }_{- 35.82 }  $ \\ 
\enddata  
\label{tab:lag} 
\end{deluxetable}